\documentclass[%
 aps,
 amsmath,amssymb,
 aps,
]{revtex4-2}

\usepackage{CJK}
\usepackage{graphicx}
\usepackage{dcolumn}
\usepackage{bm}

\usepackage{tabularx} 
\usepackage{amsmath}  
\usepackage{graphicx} 
\usepackage{natbib}
\usepackage{amstext}
\usepackage{amssymb}
\usepackage{amsmath}
\usepackage{booktabs}
\usepackage[caption=false]{subfig}
\usepackage[normalem]{ulem}

\usepackage[final]{hyperref} 
\hypersetup{
	colorlinks=true,       
	linkcolor=blue,        
	citecolor=blue,        
	filecolor=magenta,     
	urlcolor=blue         
}
\usepackage{color, colortbl}
\usepackage{comment}

\usepackage{tikz}

\usepackage{tikz-3dplot}
\usetikzlibrary{3d}
\usepackage[skins]{tcolorbox}
\definecolor{dgreen}{RGB}{63,127,0}
\definecolor{dred}{RGB}{144,14,3}
\definecolor{LightCyan}{rgb}{0.88,1,1}

\definecolor{GMcolor}{rgb}{0.1,0.1,1}
\definecolor{PAcolor}{rgb}{0.1,0.8,0.1}
\definecolor{FBcolor}{rgb}{1,0.1,0.1} 
\definecolor{PBcolor}{rgb}{1,0,1}

\usepackage{blindtext}
\usepackage{amsthm}

\newcommand{\X}{\mathbf{X}}
\newcommand{\x}{\boldsymbol{x}}

\begin{document}

\preprint{APS/123-QED}

\title{Probabilistic forecasts of extreme heatwaves using convolutional neural networks in a regime of lack of data}
\author{George Miloshevich, Bastien Cozian, Patrice Abry, Pierre Borgnat, and Freddy Bouchet}
 \email{Freddy.Bouchet@cnrs.fr}
 \affiliation{%
ENSL, CNRS, Laboratoire de Physique, F-69342 Lyon, France
}%

%
%


\date{\today}

\begin{abstract} 
Understanding extreme events and their probability is key for the study of climate change impacts, risk assessment, adaptation, and the protection of living beings. Extreme heatwaves are, and likely will be in the future, among the deadliest weather events. Forecasting their occurrence probability a few days, weeks, or months in advance is a primary challenge for risk assessment and attribution, but also for fundamental studies about processes, dataset and model validation, and climate change studies. In this work we develop a methodology to build forecasting models which are based on convolutional neural networks, trained on extremely long 8,000-year climate model outputs. This approach is parallel to weather model forecasting and has complementary scopes. Because the relation between extreme events is intrinsically probabilistic, we emphasize probabilistic forecast and validation. We demonstrate that neural networks have positive predictive skills, with respect to random climatological forecasts, for the occurrence of long-lasting 14-day heatwaves over France, up to 15 days ahead of time for fast dynamical drivers (500 hPa geopotential height fields), and also at much longer lead times for slow physical drivers (soil moisture). This forecast is made seamlessly in time and space, for fast hemispheric and slow local drivers. The method is easily implemented and versatile. We find that the neural network selects extreme heatwaves associated with a North-Hemisphere wavenumber-3 pattern. We argue that this machine learning approach should be key in the future for quantitative process studies, model intercomparisons, and dataset studies. For instance, we find that the 2 meter temperature field does not contain any new useful statistical information for heatwave forecast, when added to the 500 hPa geopotential height and soil moisture fields. The main scientific message is that most of the times, training neural networks for predicting extreme heatwaves occurs in a regime of lack of data. We suggest that this is likely to be the case for most other applications to large scale atmosphere and climate phenomena. Depending on the information to be learned, training might require dataset lengths as long as several thousands of years, or even more, for optimal forecasting skill. For instance, using one hundred years-long training sets, a regime of drastic lack of data, leads to severely lower predictive skills and general inability to extract useful information available in the 500 hPa geopotential height field at a hemispheric scale in contrast to the dataset of several thousand years long. Even with several thousand years-long datasets, no convergence is observed in the predictive skills coming from hemispheric geopotential height fields. We discuss perspectives for dealing with the lack of data regime, for instance rare event simulations and how transfer learning may play a role in this latter task.
\end{abstract}

\maketitle

\section{Introduction}

\subsection{Context: the need for probabilistic forecast of extreme climate events}

\label{sec:intro_climat}

\paragraph{Lack of data for the most impactful climate extremes.} 

Climate change is one of the major challenges of modern societies~\cite{IPCC_2021}, and will significantly affect humans and other living beings. Its most severe impacts are caused by rare and extreme events~\cite{IPCC_2021_extremes}. For instance, since 1998, most of the deaths which were caused by major related disasters have been linked to only three climate events~\cite{unisdr2018review}: the Western European extreme heatwave during the summer 2003~\cite{Herrera2010}, the storm surge related to cyclone Nargis in Myanmar in 2008~\cite{fritz2009cyclone}, and the extreme heatwave in Russia during the summer 2010~\cite{barriopedro2011hot,Otto_2012}, with death tolls of about $70,000$, $150,000$ and $100,000$ respectively. Each of the physical events causing these impacts where unprecedented in the historical record, in their category, as was the case for 2021 North-Western North America heatwave~\cite{philip2021rapid}.

These examples illustrate the need to study very rare events, most of them unprecedented. Faced with this scientific challenge, given the drastic lack of historical data, any statistical approach based solely on observation data is bound to fail. The only sensible approach is thus to use climate or weather model data, 
{whose} biases are properly characterized~\cite{woolings18} through process studies. We choose extreme heatwaves as our topic  because they will be among the most impactful climate extreme events in the future~\cite{IPCC_2021_extremes}, and because climate models are known to reproduce better their dynamics than other extreme events, because they are large scale phenomena less affected by small scale physics. In the present study, we will use 8,000-year long PlaSim (Planet Simulator~\cite{fraedrich2005planet}) climate model simulations in order to devise a forecast tool for extreme heatwaves in midlatitude dynamics. We will use modern machine learning techniques, as well as tune and develop them to specifically study very 
rare events.

\paragraph{The compound effects of geostrophic turbulence and slow drivers for extreme heatwaves.} 
\begin{figure}
	\centering
	\subfloat[\label{fig:jetstream}]{%
		\centering
		\includegraphics[width=0.53\textwidth]{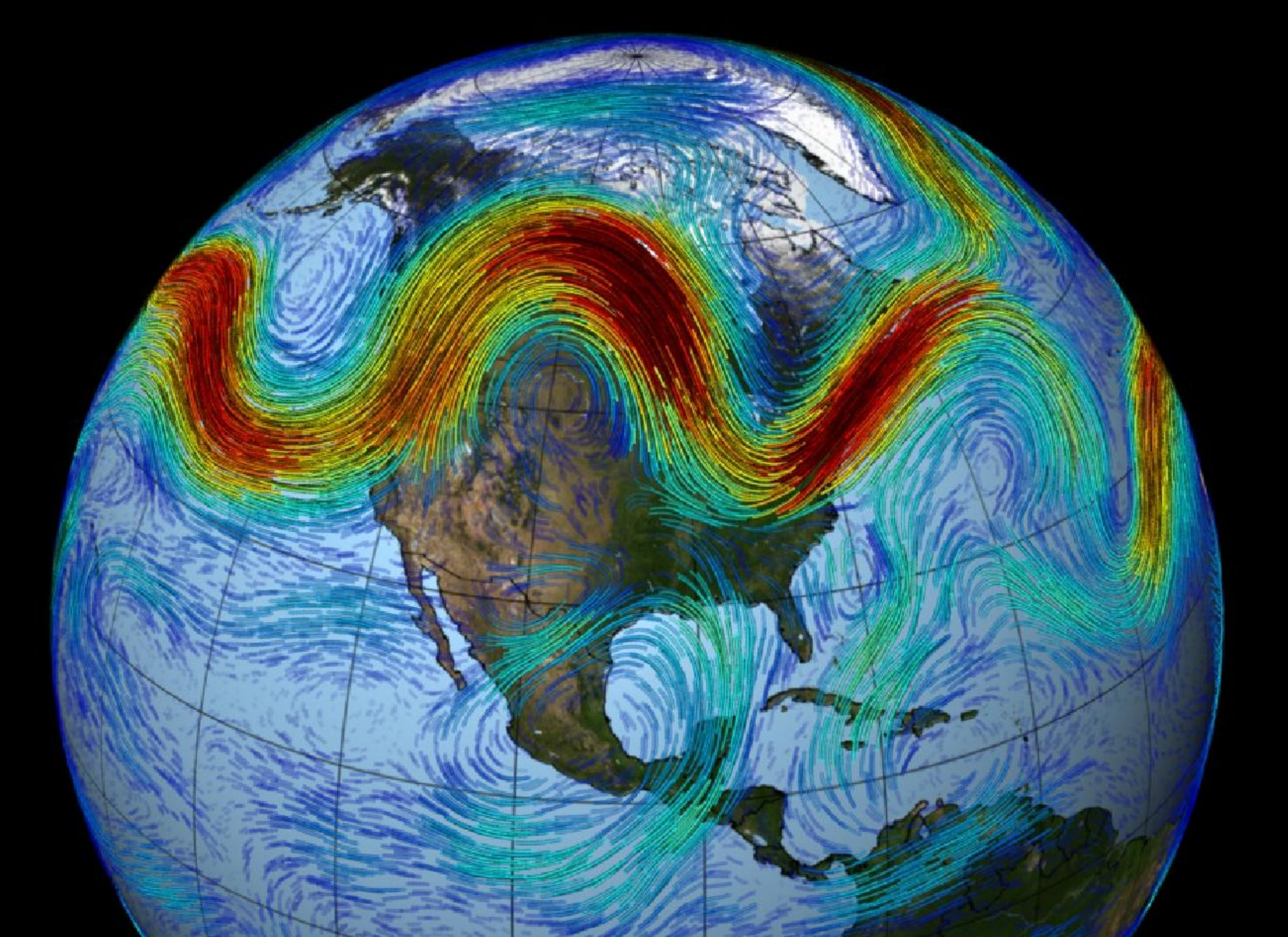}
	}\hfill
	\subfloat[\label{fig:kineticenergy}]{%
		\centering
		\includegraphics[width=0.35\textwidth]{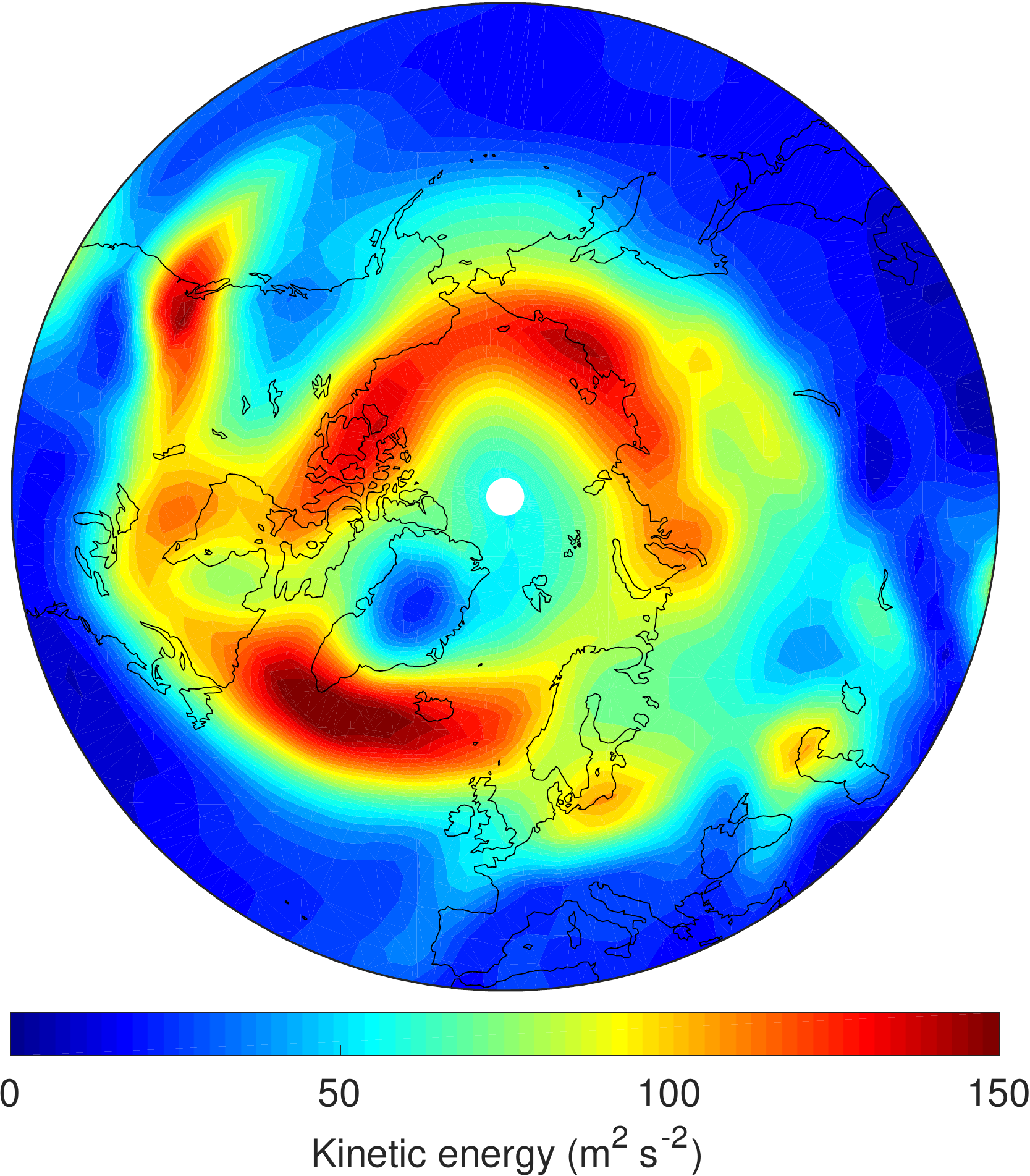}
	}
	\caption{(a) Snapshot of wind speed velocity at the top of the troposphere,
		showing the jet stream over North America (from NASA). (b) Averaged
		horizontal kinetic energy at 500hPa (mid troposphere) in the PlaSim
		model~\cite{Ragone18}, showing the averaged Northern Hemisphere jet stream}
	\label{fig:Fig1}
\end{figure}

From a fluid dynamics point of view, studying midlatitude extreme heatwaves amounts to quantifying the probabilities of rare fluctuations of the dynamics of the turbulent Earth troposphere. Midlatitude atmospheric flow is turbulent and characterized by balance between the Coriolis force and pressure gradient (geostrophic turbulence), and whose dynamics is dominated by large scale 
unstable patterns. The main large scale features are the two jet streams (one
per hemisphere). These are strong and narrow eastward air
currents, located at midlatitudes with maximum velocity of the order of $40\,\text{m.\ensuremath{s^{-1}}}$ close to the tropopause (see Figure \ref{fig:Fig1}(a)). The climatological
position of the northern hemisphere jet stream in the PlaSim model is seen on Figure~\ref{fig:Fig1}(b), that represents the time average of the kinetic energy due to the
horizontal component of the velocity field at 500 hPa pressure surfaces. The jet stream's meandering dynamics, due to non-linear Rossby waves, is related to the succession of anticyclonic and cyclonic anomalies which characterize weather at midlatitudes. It is well known that midlatitude heatwaves, like the 2003 Western European heatwave or the 2010 Russian heatwaves, are due to rare and persistent anticyclonic anomalies (or fluctuations), that arise as either blocking situations~\cite{nakamura2018atmospheric,wang2019evidence} (omega shape quasi-stationary patterns) or Rossby wave breaking or shifts of the jet stream, or more complex dynamical events leading to some quasi-stationary patterns of the jet stream. Studying extreme heatwaves then amounts to studying the non-linear and turbulent dynamics of the atmosphere, or the consequences thereof. 

Studying extreme heatwaves is however not just a problem in fluid mechanics pertaining to the extremes of turbulent fluctuations. While there is a clear connection between the physical hazard (the temperature) and the fast dynamical drivers through the fluid dynamics of midlatitude troposphere (jet stream, Rossby waves dynamics and blocking situations), it is also well known that slow drivers, sometimes also called modulators, influence the frequency and the probability of the fast dynamical drivers~\cite{horton2016review,perkins15}. For instance, deficit of soil moisture acts as a positive feedback on heatwave situations~\cite{shukla1982influence,rowntree1983simulation,dandrea06,vautard2007,fischer2007soil,lorenz2010persistence,Stefanon_2012,hirschi2011observational,schubert2014northern,zhou2019land,benson2021characterizing,zeppetello2022physics,vargas2020projected}. Indeed, in normal conditions, evaporation of soil moisture cools the ground and partition the heat flux from the heated soil to the low altitude air masses into latent and sensible heat. This effectively cools the lowest part of the atmosphere. The related produced cloud covers might also in complement affect the radiative balance. By contrast, lack or low values of soil moisture, will thus favor hotter heatwaves, while persistent heatwaves themselves or precipitation deficit might favor low soil moisture values (see~\cite{horton2016review,zeppetello2022physics} for a more precise discussion of the related physical phenomena). This leads to a strong soil-moisture deficit/heatwave positive feedback. Soil moisture is called a slow driver, or modulator, because the typical time scales of its variations, from weeks to seasons, is much longer than the synoptic timescales associated with midlatitude turbulent dynamics, of a few days up to a week. Beyond these land/atmosphere couplings through soil moisture, other slow drivers for midlatitude heatwaves are classically studied in the climate literature~\cite{horton2016review,perkins15}, for instance sea surface temperature at ocean scales, tropical or stratospheric forcing related to low modes of variability of the climate system (ENSO, QBO, and so on). It has been identified for a very long time that most extreme events are compound events of several drivers, and a recent work proposes a qualitative classification of compound types~\cite{zscheischler2020typology}. 

Understanding the relative effect of fast dynamical drivers and slow physical drivers is very interesting from the fundamental fluid mechanics perspective: it amounts to understanding the effect of slow varying and weak boundary condition changes on turbulent statistics. It is also critical in order to predict the impact of climate change on future heatwave probabilities~\cite{field2012managing,Berg15}. The topic of physical mechanisms behind extreme heat events~\cite{teng2012zonal,teng2013probability,branstator2017tropospheric,kornhuber2017summertime}, and how these mechanisms may change with climate change, is an emergent area of research, with much evidence still required. For instance, the dependence between temperature and precipitation is projected to increase in many land areas, particularly in the Northern Hemisphere, leading to a doubling in probability of extremely hot and dry summers on top of long-term climate trends~\cite{zscheischler2017dependence}. Diagnosis of heat event mechanisms is critical to understanding the potential for nonlinear responses in extreme heat beyond those expected from global mean warming alone~\cite{horton2016review}.

\paragraph{The need for probabilistic forecast for rare events.} 

The main aim of this work is to develop a setup for probabilistic forecast of extreme heatwaves, based on machine learning. Forecasting extreme events may be crucial for the sake of prevention and information dissemination for limiting risks through anticipated action, which is one of our main motivations. Another important complementary goal is the understanding of the fluid mechanics and physical processes leading to heatwaves. As just explained, a scientific challenge is to disentangle the effect of fast dynamical drivers, for instance atmospheric dynamics and geostrophic turbulence, and slow physical drivers, for instance soil moisture feedbacks. There is a need for a new methodology in order to achieve this goal. A very common approach in the climate literature is to plot maps of dynamical or physical variables, conditioned on the outcome of the extreme events, called composite maps. While interesting, such composite maps inform only on the state of the system once it is known that the event actually occurred (a posteriori conditioning on the event). A much more important question is to understand which states of the systems are more likely to lead to the extreme events (a priori conditioning on the state of the system). While these two properties are related through Bayes law, composite maps alone are not useful to study the more important a priori conditioning on the state of the system. In order to predict the probability of a future extreme conditioned on the state of the system, the so-called committor function, one actually has to build a forecasting tool able to estimate this probability. 

However, this forecast task is often considered as extremely difficult because of the very large amount of data needed and some methodological difficulties. A review of quantification methodologies to disentangle preconditions of high-impact events~\cite{tilloy2019review} describes regression techniques and event compositing, and stress the needed long dataset.  The probability that an extreme event occurs, conditioned on the state of the system, is a function of the system state and is called a committor function~\cite{lucente2020machine,Lucente2022committor}. This is the proper tool to disentangle the mechanisms that lead to extreme events in a fully non-linear setup, beyond usual restrictive assumptions. One of the main conclusion of this work will be that machine learning and neural networks provide a way to compute committor functions, by solving a probabilistic forecast problem. This, however, requires to understand predictions using  neural network in a probabilistic framework, as is further explained in the following sections. 

Another reason why the forecast should be probabilistic is because for turbulent flows, like the atmosphere, the relation between between meteorological fields (predictors) and extreme events is probabilistic. This is for three reasons. First, as originally understood by Lorenz in his 1969 paper~\cite{lorenz1969predictability}, for many chaotic dynamical systems with many degrees of freedom and a hierarchy of spatial and temporal scales, the memory of the initial condition is lost after a finite amount of time and the dynamics behave in an intrinsically stochastic way. In the case of Earth's atmosphere, the prediction of synoptic scales is intrinsically stochastic after a few days to a week~\cite{lorenz1969predictability}. Second, because for practical reason we have an incomplete knowledge of the initial conditions, then the initial condition should be considered as stochastic. Third, because the predictors we use do not describe the complete set of initial conditions. This means that assuming a one-to-one relation between the predictors (physical fields) and prediction (extreme heatwave after a $\tau$-day delay) does not make sense, even in principle. The relation between predictors and prediction should be probabilistic. Our task will actually consist in predicting the occurence of extreme heatwaves starting $\tau$ days ahead, given the knowledge of some physical fields that characterize the state of the atmosphere and soil moisture today. This is actually a classification task: given some images, or data stored in a vector, one seeks to associate a class among two: either the heatwave occurs (class one) or not (class two). From the point of view of machine learning this is very similar to image recognition. However, when recognizing the presence of a cat in an image, a one-to-one relation between the image and the class actually exists: either a cat is present on the image or not. The machine learning tool can then associate a probability to the prediction, which can be interpreted as the level of confidence of the tool due to its practical limitations, associated for instance with incomplete training or lack of data. By contrast, when predicting extreme events for a chaotic dynamical system, the relation between the predictors and the classes is intrinsically probabilistic. Then the probability given by the forecasting tool should be interpreted as intrinsic, and reflect both intrinsic uncertainty due to the unknown real probability, and practical uncertainty due to the limitations of the learning tool. We will see the consequence of this remark on the machine learning implementation and testing.

\subsection{State of the art for machine learning approaches for forecasting climate extremes}
\label{sec:intro_machine_learning}

Extreme event prediction and, more broadly, weather forecast, have recently attracted numerous studies which exploit machine learning techniques. This is contrasted with the mainstream approach which involves running expensive numerical weather prediction models. This dichotomy between physics based and pattern based prediction is well documented in the review articles \cite{balaji2021climbing},~\cite{reichstein19}. Notably there are some studies which attempt to bridge the gap by combining the approaches~\cite{karpatne17}. Overall pattern-based techniques such as neural networks or analog method may do relatively well in seasonal/sub-seasonal forecasting~\cite{cohen19}, at time scales longer than Lyapunov time. 

As an example, a forecast tool of the El-Nino Southern Oscillation (ENSO) index has been built using a Convolution Neural Network (CNN)~\cite{Ham}. A model pre-trained on CMIP ensemble was then trained on historical reanalysis. Other developments include~\cite{weyn19} where 500 hPa geopotential height was predicted using gridded reanalysis data. 

Deep learning has been applied to the spatial and temporal detection of extreme weather events such as hurricanes~\cite{liu2016application,recah17}, tropical cyclones~\cite{roisin20}, droughts~\cite{agana17,dikshit20}, storm surges~\cite{Chen22} and wind power generation~\cite{peng21}, and heatwaves~\cite{Chattopadhyay19,jacques-dumas22}.  
For further reference see~\cite{mudigonda21} and the citations therein. Recently, for predicting extreme heat events globally, neural networks trained on reanalysis data~\cite{lopez2022global} have given positive skill compared to the ECMWF subseasonal-to-seasonal control forecast after two weeks. Neural networks, where the 500 hPa geopotential height and surface temperature were used as predictors, were able to predict both short duration~\cite{Chattopadhyay19} or long-lasting~\cite{jacques-dumas22} heatwaves, when trained on climate model data. In these works, the performance was evaluated and tuned to metrics related to confusion matrix such as Matthew's Correlation Coefficient, which are well suited for one-to-one or deterministic relation between predictors and prediction. However making and testing probabilistic forecast is very important, as stressed in section \ref{sec:intro_climat}. Changing this paradigm requires to  test probabilistic forecasts using probabilistic scores. Such probabilistic scores have been used for a long time in evaluation of weather model forecast, for instance the logarithmic score which is an objective and proper score~\cite{benedetti10}.  Another proper score, although less applicable to rare events, a Brier score, was used in a recent study~\cite{straaten22} where subseasonal forecast was made for high temperatures in western and central Europe using random forest approach applied to ERA5 reanalysis. In general, traditional techniques such as random forest are quite competitive with  neural network approaches when dealing with smaller dataset sizes. Finally, we stress the work by~\cite{delaunay22} that produces state-dependent probabilistic Madden-Julian oscillation forecast with neural networks. 

One of the aims of this work will be to perform and test \emph{probabilistic} forecasts for the first time to the best of our knowledge, using  neural networks, for extreme climate events or atmospheric dynamics phenomena. Probabilistic forecast will be performed through a natural interpretation of Softmax probabilities. When working with rare events, because of learning difficulties with class imbalance, it might be useful and efficient to undersample the majority class. Using such undersampling at the same time as making a probabilistic forecast however requires an interpretation of Softmax probabilities that takes into account the undersampling rate~\cite{fernandez2018learning,pozzolo15a}, as  will be explained.

In a recent work~\cite{jacques-dumas22},  neural networks have been used in order to forecast extreme heatwaves. One of the key originality of this first work was to consider for the first time the forecast of long-lasting extreme heatwaves, with durations of several weeks. This is a key point as most of the extreme heatwaves with the largest impact, for instance the Western European one in 2003, the Russian one in 2010, or the North American Pacific coast one in 2021, lasted long, from two to five weeks. The lack of comprehensive studies of the statistics of long-lasting events has actually been stressed in the last IPCC report \cite{IPCC_2021}. We refer to the introduction and section 2.1 of~\cite{jacques-dumas22} for a thorough discussion of this crucial point about the definition of extreme heatwaves. Other key achievements of~\cite{jacques-dumas22} were to demonstrate the efficiency of  neural network to predict long-lasting events, to implement and assess the interest of large-class undersampling and transfer learning. From the point of view of machine learning methodology, this new paper builds on the previous one~\cite{jacques-dumas22}, but with several crucial methodological improvements: probabilistic forecasts and tests, implementation of large class undersampling in a probabilistic setup, and use of both fast and slow physical and dynamical drivers. Another distinction in this new paper is that we work with fields in the physical space rather than in the Fourier space. This proves more efficient from the point of view of the forecast skills, and especially so when studying the importance of local versus global information for best performance. We also use a much longer, 8,000-year dataset, which represents the climate of the decade 1990-2000 with a more realistic daily cycle, and which allows for a detailed study of the lack of data regime and a more comprehensive analysis of the various drivers. 


\subsection{Goals, contributions and outline} 
\label{sec:intro_goals_contribution}

Section \ref{sec:intro_climat} discusses the importance of forecasting long-lasting extreme heatwaves because of their impact. We have also reviewed the large interest in the climate literature for understanding the respective effects of fast dynamical drivers, related to troposphere dynamics, geopotential height and temperature maps, and slow physical drivers, for instance soil moisture. We have stressed that it is crucial that this forecast should be probabilistic and that it has to be performed in a regime of lack of data. 

In order to build a machine learning forecast setup that will be able to address these goals,  several new methodological contributions are proposed in this work. To devise and use  neural networks that predict probabilistic forecasts will be our first methodological goal. The neural network output will be the probability of the extreme event, as a function of the state of the system, also called a committor function. As a second methodological goal, this probabilistic forecast will be validated using a probabilistic score. Because of the regime of lack of data and large class-imbalance, we will propose and test a large class undersampling strategy adapted to probabilistic forecast, as a third methodological goal. We will demonstrate the efficiency of these three methodological contributions for predicting long-lasting extreme heatwaves using climate model outputs.  

Using this  neural network technique, adapted and validated in a probabilistic framework, the following fluid mechanics and climate goals will be addressed. First the prediction capabilities of the neural network when changing the predictor fields will be investigated. We will demonstrate that the network is able to make best predictions when combining fast and slow drivers, with a relatively stronger contribution of fast drivers for shorter lead times and stronger contribution of slow drivers for longer lead times. Second, the effects of dataset lengths will be studied , a very important question in a regime of lack of data. We will actually conclude that the dataset length has to be extremely long for proper convergence of the learning, and that in such a regime, the optimal learning results in a tradeoff between the size of the physical domain and data availability. For instance, for predicting extreme heatwaves over France, it is optimal to use local data (North Atlantic and Europe) with one hundred-year long datasets, while it is optimal to use global data for few thousand-year long datasets. Finally, in order to make a connection with fluid-mechanics, we will study the interpretability of the learned committor function by computing composite maps conditioned on high extreme event probabilities. 

Those goals will be studied using long datasets from the Planet Simulator (PlaSim) climate model \cite{fraedrich2005planet,fraedrich_1998}. This model has a very realistic fluid dynamics component, similar to the climate models used for CMIP experiment described in IPCC reports. However its physical parameterizations are simpler than such models, for instance the ones used for CMIP6 experiments. This significantly reduces PlaSim's computation time by about a factor 100. It is thus suitable for methodological development and first studies, using extremely long datasets. It is ideal for studying learning convergence in the lack of data regime. Section~\ref{sec:data} gives a more detailed introduction to the PlaSim model, its output fields, and the dataset size, resolution, quality and richness. It also discusses better the physical interest and limitations of this dataset. In section~\ref{sec:data}, heatwaves, predictor fields, and the probabilistic prediction problem studied in this work are also defined precisely. 

Section~\ref{sec:predprob} formalizes the problem of probabilistic forecast using  neural networks, and discusses proper probabilistic scores and their relationship with cross-entropy and machine learning loss functions. We also introduce a very useful Normalized Logarithmic Score which is positively oriented (the larger the better), and takes value zero for prediction according to the climatological frequency and one for perfect prediction. In addition, the relation of these scores to the Brier score, another classical probabilistic score, is  discussed.

Section~\ref{sec:NN} 
first describes the Convolutional Neural Network architecture used here and its learning conditions. It further explains the promoted probabilistic strategy:
 to train the architecture as a classifier and to use it as a conditional probability predictor. The detailed training protocol using a classical cross-validation procedure, to assess confidence and reproducibility in performance, is presented. Finally, it explains the methodology for large class undersampling in a probabilistic framework.

Section~\ref{sec:results} presents the fluid mechanics and climate science results. It first demonstrates which combination of physical predictors, among slow and fast drivers, give the best prediction skill. This stresses the potential of machine learning for dealing adequately with separating their respective effects. We then discuss the importance of the dataset length, the convergence of the learning skill when the dataset length is changed, and the tradeoff between dataset length and spatial extension of the physical fields. We verified that our neural network has a better performance than traditional approaches, for instance logistic regression using Empirical Orthogonal Function (EOF) decomposition (or PCA), although the details of this analysis are not reported in this paper. We also test the continuity and consistency of the committor function prediction when the time lag is varied. 

Finally, section~\ref{sec:conclusion} discusses conclusions and perspectives.



\section{Long-lasting heatwaves and Planet Simulator data}
\label{sec:data}

Using weather maps, we aim at predicting the probability of occurrence of extreme long-lasting heatwave that starts $\tau$ days later. In the remainder we will refer to this parameter as \emph{lead time} or sometimes \emph{lag time}. We first define long-lasting heatwaves in section \ref{sec:defheatwave}, describe the possible predictors in section \ref{sec:InputData}. In section \ref{sec:committor}, we explain that our approach is actually a way to compute the committor function, a key function in the field of rare event analysis and simulations. The actual weather maps we use are outputs of the PlaSim climate model, which is described in details in section \ref{sec:plasimdata}.

\subsection{What are long-lasting heatwaves?}\label{sec:defheatwave}

Several indices have been used in the meteorology, climate, and impact literature, to define heatwaves, for different purposes \cite{perkins15}. However, long-lasting heatwaves are the most detrimental to health \cite{barriopedro2011hot} and other living beings.  Moreover, many of the extreme heatwaves with the largest impact, for instance the Western European one in 2003, the Russian one in 2010, or the North American Pacific coast one in 2021, lasted long, from two to five weeks. They were often composed of several sub-events with the classical definitions \cite{perkins15}. We want to use a definition of heatwaves that actually involves a measure related to both the persistence and the amplitude of air temperature close to the ground.

We thus define heatwave as time and area average of daily 2-meter temperature. Seminal studies \cite{schar,barriopedro2011hot,coumou12} of the 2003 and 2010 heatwaves already considered averaged temperature over variable long time periods (7 days, 15 days, one month, three months). Several recent works~\cite{Ragone18, galfi2019, Ragone19, galfi21, Ragone21, galfi21} have studied heatwaves based on time and space average of either the 2-meter temperature or of the surface temperature. This viewpoint is expected to be complementary with the classical definitions \cite{perkins15}, and quite relevant to events with the most severe impacts.  Such definitions have the advantage to define events at a synoptic scale which are geographically located and which begin at a specific date. This is well suited for a forecast perspective. 

\begin{figure}
	\centering
	\centering
	\includegraphics[width=8.6 cm]{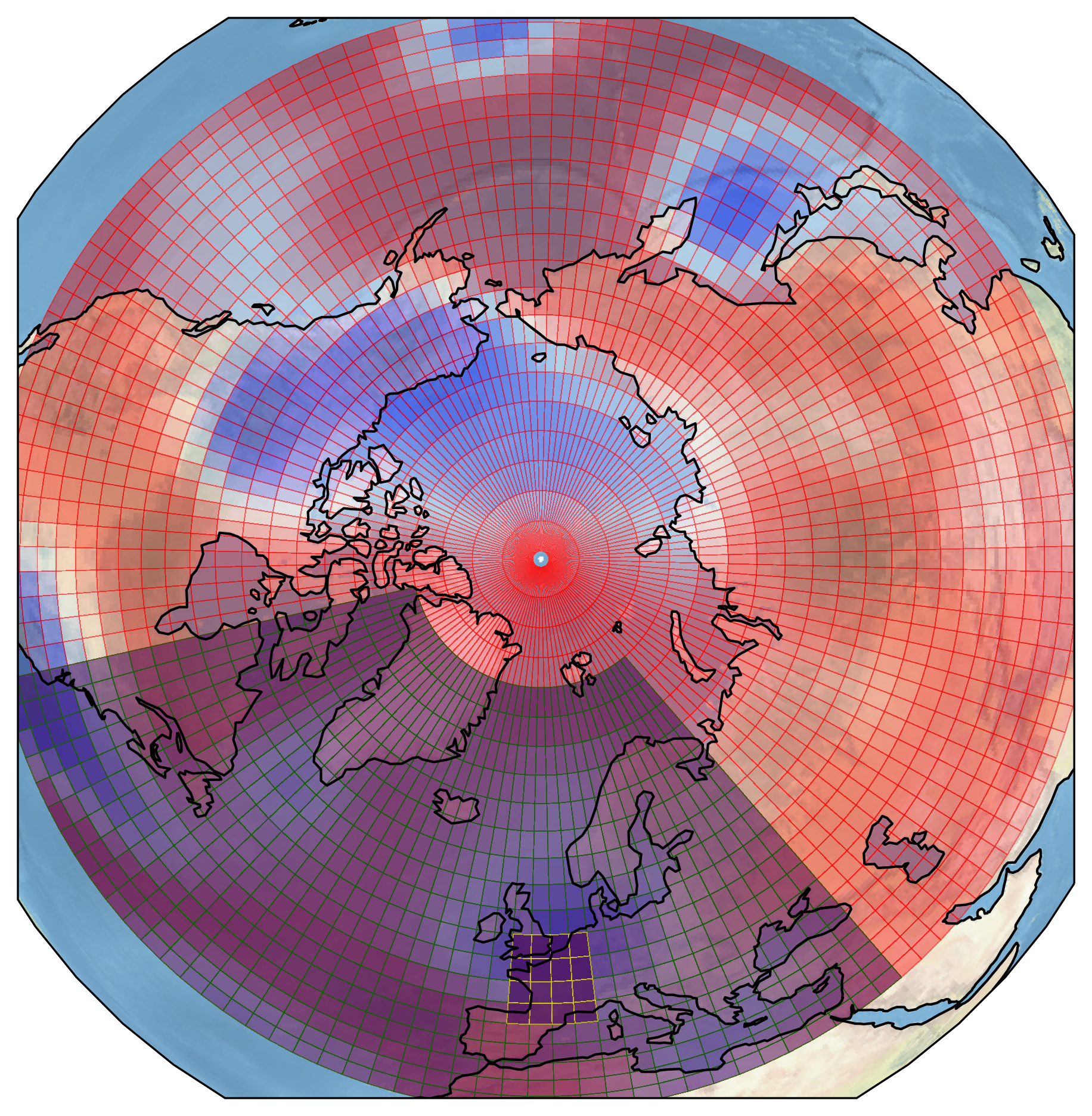}
	\caption{For all fields, we will use gridded data on the mid and high latitude Northern Hemisphere as represented by red meshlines. The figure also features in purple the area $\cal{D}$ (France). The North Atlantic Europe sector is represented in blue.} 
	\label{fig:NorthHemisphere}
\end{figure}

The daily 2-meter temperature $T_{2m}(\vec{r},t)$ is a spatial field that depends on the location $\vec{r}$ and calendar day $t$ (also called time). We use daily averages. Statistics of $T_{2m}(\vec{r},t)$ are affected by the seasonal cycle. We compute anomalies (i.e. fluctuations) by subtracting the statistical average $\mathbb{E}\left(T_{2m}\right)(\vec{r},t)$ at each point $\vec{r}$ and each time $t$. We compute the space and time averaged 2-meter temperature anomalies:

\begin{equation}\label{timeaveraged}
	A(t)
	=
	\frac{1}{T}\int_{t}^{t+T}\frac{1}{\mathcal{\left|D\right|}}\int_{\mathcal{D}}\left(T_{2m}-\mathbb{E}\left(T_{2m}\right)\right)(\vec{r},t^\prime)\,\mathrm{d}\vec{r}\,\mathrm{d}t^\prime
\end{equation}
where $T$ is the length of the time average, i.e. heatwave duration, and $\mathcal{D}$ is the spatial area for the heatwave of interest. The heatwave duration $T$ and area $\mathcal{D}$ should be understood as parameters that can be changed from one study to another, depending on the kind of heatwave of interest. $T$ ranges typically from one-day (short duration) to three months (a season). $\mathcal{D}$ is typically of the size of the synoptic scale. The synoptic scale, of about $1,000$ km at midlatitude, is the order of magnitude of correlation length for troposphere dynamics, and corresponds to the typical size of anticyclones, cyclones and the jet stream meanders. In the present work, mainly aimed at methodological developments, we set $T=14$ days (two-week heatwaves), and $\mathcal{D}$ to be the France area: the set of grid-points corresponding to France area is visible on Figure \ref{fig:NorthHemisphere}.

We consider summer statistics, during the months of June, July and August. More precisely we consider $A(t)$ for $t$ between June 1st and August 16th (inclusive), such that the time average in (\ref{timeaveraged}) involves only days during the months of June, July and August. In PlaSim each month spans 30 days, thus the total length of the period of interest is 77 days. The statistics of $A$ is considered as approximately stationary during the summer, although there is actually some very small non-stationarity. For instance, monthly breakdown for standard deviation is $\sigma_A = 1.58 {\rm K}$ in June, $\sigma_A = 1.49 {\rm K}$ in July and $\sigma_A = 1.32 {\rm K}$ in August. We see that the variations of the standard deviations from one month to another are much smaller than the standard deviations themselves, and much smaller than the variations of the time averaged temperature from one month to another.

Extreme heatwaves are defined as rare large values of the time and space average $A(t)$. Following the previous works~\cite{jacques-dumas22}, we define an extreme heatwave as an event (a day) for which the time and space averaged 2-meter temperature anomaly exceeds the threshold $\alpha$: $A(t)>\alpha$. We introduce an indicator variable $Y(t)$ which is equal to $1$ when $A(t)>\alpha$ and $0$ otherwise. We have $K=2$ classes of events: heatwaves when $Y=1$ and no-heatwave when $Y=0$. $Y(t)$ is sampled daily. When $Y(t)=1$, $t$ is the day for the start of the heatwave and the heatwave lasts for $T$ days, by definition. The threshold $\alpha$ can be changed depending on the heatwaves of interest. The number of classes $K$ could also be changed. For this methodological study, we use $K=2$ and $\alpha$ such that the heatwave class contains $5\%$ of the total number of summer days (excluding the last two weeks for the reasons explained). For the PlaSim model data described below, this corresponds to $\alpha= 2.7 {\rm K}$.

\subsection{Predictors for heatwaves}\label{sec:InputData}
\label{sec:inputs}

\begin{figure}
	\centering
	\includegraphics[width=15.2 cm]{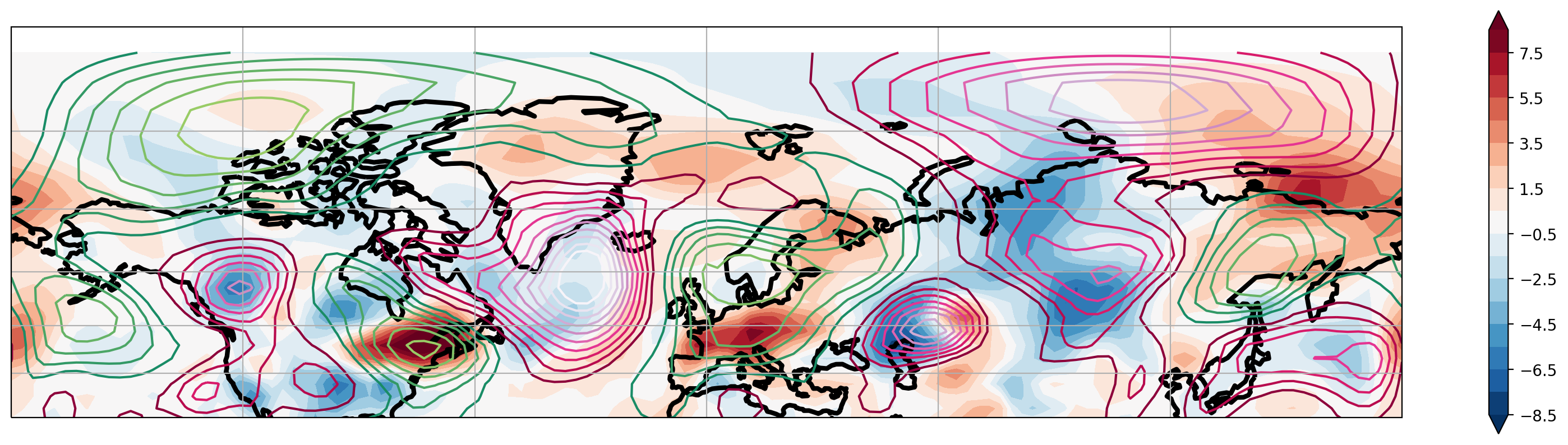}
	\caption{Snapshot of 2 meter temperature anomaly and 500 hPa geopotential height anomalies. Temperature colormap is shown on the colorbar. The geopotential height anomalies are plotted via contour lines, colored green for positive and purple for negative with a separation of 20 meters between the lines. The lowest level also corresponds to 20 meters. This map displays a synoptic situation (daily averaged) on the first day of the strongest heatwave in the dataset.}
	\label{fig:synoptic}
\end{figure}

Our objective is to develop a prediction tool for extreme heatwaves. From the knowledge of observed weather fields (predictors), we want to predict the probability of the event $\tau$ days later. We will vary the parameter $\tau$ in order to understand how the predictability potential changes with this lead time $\tau$. 

The choice of good predictors, among all possible weather fields, is a key practical and physical question. From the common knowledge among weather and climate scientists, it is known that maps of the 2-meter temperature $T_{2m}$ and of the geopotential height $Z$ (in meters), for instance on the 500 hPa iso-pressure surfaces (the 500 hPa geopotential height) are relevant variables. The geopotential height at 500hPa (close to the middle of the troposphere) is considered an excellent representation of the dynamical state of the atmosphere. Indeed, it is closely related to pressure variations at a fixed altitude, to anticyclones (positive values), and to cyclones (negative values), in the lower troposphere. Moreover, on those surfaces the wind flows along the isolines of the geopotential height, to a good approximation. The 2-meter temperature $T_{2m}$ used as a predictor is directly related to the kernel of the integral (equation~\ref{timeaveraged}) whose fluctuations we seek to predict, and gives further information on lower atmosphere dynamical processes compared to the 500 hPa geopotential height. We stress that $A(t+\tau)$ involves the time average over $T$ days of $T_{2m}(t)$. The knowledge of $T_{2m}(t)$ thus provides only partial estimate of $A(t+\tau)$, just by virtue of persistence prediction or low-tropospheric advection. However given that the correlation time is of order of a few days, to be compared to $T=14$ days, this information gives a relatively small predictive power by itself even for $\tau=0$, which quickly diminishes for $\tau$ larger than a few days. In the following, we never use directly the 2-m temperature or 500 hPa geopotential height fields, but rather their anomalies by subtracting their seasonal average. Representative examples of snapshots of 2-m temperature or 500 hPa geopotential height fields is shown on Figure \ref{fig:synoptic}. 

The 2-meter temperature $T_{2m}$ and the 500 hPa geopotential height $Z$ are fields that evolve through the chaotic dynamics of the atmosphere with a typical time called the synoptic time scale, of order of a few days. It is known that both temperature and geopotential lose the memory of the initial condition of the atmosphere, and that their auto- and cross-correlations decay, after times of order one to two weeks. This is the predictability margin for weather. As a consequence, we expect these fast dynamical fields to lose their predictive power after times of order 15 days at most. Those fields have actually been used as predictors for machine learning approaches in past studies, either for 5 day~\cite{Chattopadhyay19} or 14 day~\cite{jacques-dumas22} heatwaves. Those works have indeed demonstrated the predictive value of these fields for time delays $\tau$ up to about 15 days.\\

One of the aims of this work is to combine predictors based on fast dynamical weather fields, just discussed, with other drivers with a much slower typical evolution. As explained in the introduction, soil moisture deficits and heatwaves are coupled through positive feedback loops and reinforce each other on various time scales~\cite{shukla1982influence,rowntree1983simulation,dandrea06,vautard2007,fischer2007soil,lorenz2010persistence,Stefanon_2012,hirschi2011observational,schubert2014northern,zhou2019land,benson2021characterizing,zeppetello2022physics}. Because soil moisture is the stock of all water in the soil, it evolves on rather long time scales. Its value is correlated over weeks to months~\cite{shukla1982influence,huang1993monthly}. At a specific time, the effect of soil moisture on dynamical variables which directly cause heatwaves, is rather weak. It is basically only able to modulate the energy budget and temperature. For this reason, we expect the soil moisture predictor to have a much smaller predictive value than fast dynamical fields in the short run, but by contrast it should provide extended memory effect. One of the aims of this paper will be to test these simple qualitative ideas and to make them precise and quantitative, using the machine learning tool.

Other slow evolving drivers might be considered as good predictors for heatwaves, i.e. sea surface temperature, or slow modes of variability of the atmosphere, ice and snow cover, and so on~\cite{straaten22}. However, as explained in the next section, the dataset we use is not suited for this, and such studies will be postponed for future works.

For simplicity, in the following, we denote $Z$ the 500 mbar geopotential height, $T_{2m}$ the 2-meter temperature and $S$ the soil moisture fields.\\

A second question pertaining to the predictors is to decide whether to use either all their values in the Northern Hemisphere mid and high latitudes, or rather to use their values on a restricted area close to the heatwave region or some intermediate scale area.  
Because the soil moisture feedbacks are local processes, soil moisture is expected to be relevant locally, for instance around the area where the heatwave occurs, while geopotential height is expected to play a role on a more hemispheric scale. For instance for many past studies of weather and climate phenomena over Europe, values of predictors over the North Atlantic and European area were typically used for the geopotential height. As examples, several very interesting studies, using analogs as a learning tool, have shown that the North Atlantic and Europe sector might be the best choice~\cite{yiou2014anawege,yiou2019stochastic}, with the interpretation that this area carries the most relevant information. For temperature, it is less straightforward to assert a priori whether just local or hemispheric information matter for heatwave prediction.

We will use the machine learning tool to assess the question of such optimal predictors. We will use either the Northern Hemisphere mid and high latitude fields, corresponding to the values of the fields above 30N, spanning a $22 \times 128 $ grid points, or to a restricted area, corresponding to a $18 \times 42 $ grid points, referred as the North Atlantic and European sector, or to the France area (see Figure~\ref{fig:NorthHemisphere}). On these restricted areas, the 500 hPa geopotential fields $Z$, the 2-meter temperature $T_{2m}$, and the soil moisture $S$ are denoted respectively $Z_{NH}, T_{NH}, S_{NH}$ for the North Hemisphere, $Z_{NAE}, T_{NAE}, S_{NAE}$ for the North Atlantic and European sector, and   $Z_{F}, T_{F}, S_{F}$ for France (see also Section~\ref{ssec:training}). 

These data can be stacked, as different input features, for the learning procedure. It was shown in~\cite{jacques-dumas22} that stacking was the best approach for combination in this context, as it allows to capture interaction between learned features.\\

More abstractly, the set of predictors (one of the combinations of $Z_{NH}, T_{NH}, S_{NH}$, $Z_{NAE}, T_{NAE}, S_{NAE}$, or $Z_{F}, T_{F}, S_{F}$), is globally called $\X\in\mathbb{R}^{d}$. From the value of the vector $X(t-\tau)$ at some time $t-\tau$, we aim to predict the probability $p(\x)$, that $Y(t)$ is equal to one (to observe an heatwave at time $t$), given that $\X = \x$. This is a probabilistic classification task, conditioned on the state $\x$. In the context of stochastic processes $p(\x)$ is called a committor function, as we will explain below.

\subsection{Committor functions for extreme heatwaves}\label{sec:committor}

We note that in the theory of rare events for stochastic processes, the probability $p(\x)$ to observe a rare event conditioned on the state of the system $\x$, is called a committor function~\cite{onsager1938initial,weinan2005transition}. The committor function is the probability of hitting the target set $\mathcal{B}$ before the set $\mathcal{A}$: $\mathbb{P}\left(\tau^\star_{\mathcal{B}}(\x)<\tau^\star_{\mathcal{A}}(\x)\right)$, where $\tau^\star_{\mathcal{A}}$ and $\tau^\star_{\mathcal{B}}$ are the first hitting times of the sets $\mathcal{A}$ and $\mathcal{B}$, given that the trajectory started at $\x$. It is possible to extend this definition to time dependent sets $\mathcal{A}$ and $\mathcal{B}$ with an extended dynamical system, see~\cite{Lucente2022committor}. For example in our case, the set $\mathcal{A}$ is simply the set of the model fields ($\x$) such that we have a heat wave which starts at time $\tau$, and the set $\mathcal{B}$ is the complementary set to $\mathcal{A}$. For each value of $\tau$, the probability $p$ that we seek to predict is then a committor function.
		
		Committor functions are extremely useful in the simulation and prediction of rare events. Several computations of committor functions have been performed with applications in either geophysical fluid dynamics or in climate sciences~\cite{finkel2021learning,miron2021transition,finkel2020path,lucente2020machine,Lucente2022committor,Lucente2022}, using either direct or involved approaches. However, computing or sampling a committor function is a very difficult task, especially in large dimensional spaces, because it requires to gather a very large amount of statistics about rare events. 
		
		Many interesting methods have been or are currently being devised to learn committor functions: based on direct machine learning~\cite{pozun2012optimizing}, or using approximations of the stochastic dynamics, for instance using the analogue approach~\cite{Lucente2022}, or Galerkin approximations of the Koopman operator~\cite{thiede2019galerkin,strahan2021long}. The present work, by successfully implementing a  neural network that efficiently forecasts $p(\x)$ the probability of extreme heatwaves conditioned on the state of the system $\x$, demonstrates that  neural networks are useful and efficient to compute committor functions for extreme heatwaves, for a dynamics that takes place in a state space with about $10^6$ degrees of freedom.
		
		\subsection{Data from the Planet Simulator model}\label{sec:plasimdata}
		
		In this work, we will use a very long 8,000-year dataset obtained as the output of the PlaSim climate model. In this section we briefly describe the model and its specific implementation and climate for producing this dataset. We also compare it to other climate models and explain its potential limitations and why it is suited for the present study. 
		\\
		
		The Planet Simulator (PlaSim) climate model~\cite{fraedrich2005planet,fraedrich_1998} solves the global dynamics of the Earth atmosphere, coupled to ocean, ice, and land surface components. Its atmosphere dynamical core solves the primitive equations for vorticity, divergence, temperature, and pressure. The governing equations are solved using a spectral method. Unresolved processes, such as radiation, interactive clouds, moist and dry convection, large-scale precipitation, boundary layer fluxes of latent and sensible heat, and vertical and horizontal diffusion, are parametrized. The land component of the model deals with the dynamics of soil moisture, which is a key physical component of the land-atmosphere feedbacks, as long as heatwaves are concerned. It is modeled by a single-layer bucket model~\cite{manabe69}. Soil water is replenished by precipitation and snow melt and is depleted by the surface evaporation.  Soil water is limited by a field capacity with prescribed geographical distribution. If the field capacity is exceeded the runoff is provided to the river transport scheme.
		
		For computing this specific dataset, the model is set up to run with fixed greenhouse gases concentrations and boundary conditions (incoming solar radiation, sea surface temperature and sea ice cover distributions) cyclically repeated every year, in order to generate a stationary state reproducing a climate close to the one of the 1990’s~\cite{fraedrich2005planet,fraedrich_1998}. For instance, the sea surface temperature is seasonally varying along the year, a cycle which is repeated each year. The horizontal resolution is T42 in spectral space, corresponding to a spatial resolution of about $2.8$ degrees in both latitude and longitude. In practice, the horizontal fields of data have a spatial size of $64\times128$ pixels, covering the entire globe. The vertical resolution corresponds to 10 vertical layers. Moreover, each field is sampled in time at $\delta t= 3$ hours sampling period, and daily averages are taken in the analysis stage. The 8,000-year dataset is obtained by 80 runs with independent initial conditions, each 100-years long. 
		
		We have already used a similar setup in previous works~\cite{Ragone18,jacques-dumas22}. In the new 8,000-year dataset used in this work, the setup is slightly different compared to our previous 1,000-year dataset. We use a diurnal cycle, which is more realistic compared to the previous studies with diurnal variation of the solar forcing and we predict 2 meter temperature rather than surface temperature, which is more relevant for impacts and a bit more difficult to predict. We note that in this paper we use daily rather than 3 hour average for prediction in contrast to~\cite{jacques-dumas22}. In an ablative study, we have trained neural networks with the 1,000-year datasets and concluded that the predictive skills are similar with or without daily cycles, and independent of whether 3 hour long samples or daily averages were used for the prediction.\\
		
		The PlaSim model has physical parameterizations that are of a lesser quality compared to up-to-date climate models which are used for CMIP experiment, analyzed in many studies documented in the IPCC reports. Its advantage, however, is that it runs about 100 times faster, and is specifically suited for producing extremely long datasets. No other statistically stationary dataset with 8,000 years of fixed present-day climate simulation is available using climate models for CMIP experiments. The atmosphere dynamics component of the PlaSim model is equivalent to many of the CMIP models, although the forcings and couplings are slightly degraded. The obtained large scale fields and patterns are of an excellent quality \cite{fraedrich2005planet,fraedrich_1998}. In the work in preparation, we show that the composite statistics of the large scale 500 hPa geopotential height fields, conditioned on heatwaves, leads to very similar patterns for this PlaSim model dataset, for CESM model outputs for a 1,000-year similar climate, and for the ERA reanalysis dataset~\cite{ecmwf2020}. CESM is one of the best models used for CMIP experiments. Because of the physical setup, for instance the lack of an active ocean in the model simulation, given PlaSim dataset is not suited for other process studies, for instance the impact of sea surface temperature fluctuations. For the study of other slow physical drivers than soil moisture, other climate model outputs might be needed. 
		
		For all of these reasons, and because having a very long dataset was key, this PlaSim model dataset was suited for this study. This allowed us to emphasize the methodological development and the study of the training convergence with the dataset length, in the lack of data regime. 
		


\section{Prediction of probability for rare events}
\label{sec:predprob}

In this section we discuss how to make a probabilistic forecast using a  neural network, for a classification task, using Softmax probabilities. We also define and discuss the scores for probabilistic forecast, and their relation with the neural network score functions. 

\subsection{Softmax Regression for the inference of the probabilities of rare events}

\subsubsection{Setting}

The task we consider is the inference of probabilities of having events $Y\in \{(0,1)\}^K$  from a set of physical observables (or features) globally called $\X\in\mathbb{R}^{d}$. If one event is in the class $k$, then $Y_k=1$ and for $l\neq k$, $Y_l=0$ (the classes are exclusive, and each event belong to one and only one class).  

As explained in section~\ref{sec:defheatwave}, for the specific application in this paper, $K=2$, $Y=1$ when a heatwave occurs at time $t$ and $Y=0$ otherwise, and $\X$ represents all the relevant dynamical and physical predictors of the state of the atmosphere at time $t-\tau$. However the following discussion is general for any probabilistic classification and is independent on a specific dataset or network architecture. 

Let us consider the pairs $(\X,Y)$ as random variables, having a ground truth joint probability distribution 
$\text{\ensuremath{\mathbb{P}\left(\X=\x\,\,\,\text{and}\,\,\,Y=y\right)=P(\x,y)= \mathbb{P}(Y=y|\X=\x)\mathbb{P}(\X=\x)}.}$ The objective is the soft assignments of features into the event types, in the sense that we want to estimate a probability that a given realized state $\x$ will conduct to the different possible classes.

Hence, the task is the inference of the probability density  $\mathbf{p}=\left\{ p_{k}(\x)\right\} _{0\leq k\leq K-1}$
where $p_{k}(\x)=\mathbb{P}\left(Y_k=1\left|\X=\x\right.\right)$ is the probability that $Y_k=1$, given that $\X=\x$. Because the classes are exclusive, we have $\sum_k p_{k}(\x) =1$. 

We classically formulate the inference of $\hat{\mathbf{p}}$ (the estimated values of $\mathbf{p}$) as a soft classification problem, relying on using a softmax function at the end of the proposed learning architecture (see the detailed architecture in Section~\ref{NNarchitecture}). Softmax probability offers a convenient way to output, for each input $\x$, an output having the meaning of the probability instead of only a class of events.

\subsubsection{The Softmax probabilities}

Softmax parametrization is a way to output probabilities associated with a discrete variable. If taking directly the features that pass through a single (not hidden) layer as input, it is equivalent to what is known as logistic regression. For a detailed explanation, please see~\cite{MEHTA20191} for a review for physicist, or \cite{Goodfellow-et-al-2016} for a textbook in Machine Learning.
The output probability has to be a positive function that should sum up to 1 over all the $K$ classes. To force that, in logistic regression, the probability associated with the features $\boldsymbol{x}$ is modeled by first computing a vector in  $\mathbb{R}^{K}$ written as $\mathbf{o}(\boldsymbol{x},\boldsymbol{\theta})  = \mathbf{W} \boldsymbol{x}  + \mathbf{b}$, where $\boldsymbol{\theta} = (\mathbf{W}, \mathbf{b})$, and then taking the normalized exponential (also known as Softmax function) of this vector to model the probabilities:
\begin{equation}\label{ProbabilityLabelGivenWeights}
	{
		P\left(Y_k = 1 \mid \boldsymbol{x},\boldsymbol{\theta}\right)=\frac{\mathrm{e}^{-\mathbf{o}_k(\boldsymbol{x}, \boldsymbol{\theta})}}{\sum_{j=0}^{K-1} \mathrm{e}^{-\mathbf{o}_j (\boldsymbol{x}, \boldsymbol{\theta})}}.
	}
\end{equation}
The quantity $\mathbf{o}$, which is the non normalized log probability, is also called the logit.
Here, $\mathbf{o}_k$ is the output associated with the discrete variable $Y_k$, and in this situation of logistic regression, it would be simply the linear form $ \mathbf{o}_k = \mathbf{w}_k \boldsymbol{x} + \mathbf{b}$.
More generally when using neural networks (see later in Section \ref{NNarchitecture}), the logit is a non-linear function $\mathbf{o}(\boldsymbol{x},\boldsymbol{\theta})$, involving several layers with parameters $\boldsymbol{\theta})$, where the original features are input of the first layer. In that situation, the softmax function is the last layer of the neural network. 

This softmax parametrization achieves probability regression, as will be described here. Note that  it can be used also for classification if a threshold for probability counting as a positive event (heatwave) is chosen; that was the purpose of the previous work on the prediction of heatwaves in \cite{jacques-dumas22}, and the results were only discussed in terms of categorical prediction, with a focus on recall (also called sensitivity in the binary case, i.e. the True Positive rate) and False Positive rate (1 minus the specificity, i.e. the fraction of false positive among all true negative events) of events. Here, we will study with more details the obtained probabilities on the different classes. A discussion in terms of TP and FP rates, and of the commonly used Matthews Correlation Coefficient (MCC) as a metrics to combine the two~\cite{MATTHEWS1975,chicco2020advantages}, as done in~\cite{jacques-dumas22}, while useful and sometimes providing qualitatively similar results, it is not properly adapted to quantify the skill of the inference of probabilities. For this purpose probabilistic score is requested.

\subsection{Proper probabilistic score, learning loss function, and normalized logarithmic score}\label{sec:properscoreasloss}

In the meteorology and climate fields, a huge literature has been devoted to the definition of good scores for probabilistic forecast validation. A good probabilistic score should be additive with respect to new events, proper (it should be maximum when the forecasted probability is the ground truth probability), and should not depend on unobserved events~\footnote{e.g. Brier depends on unobserved events in multi-class classificaiton problem}.  In the case of probabilistic classification, of interest in this paper, we follow the analysis of Benedetti~\cite{benedetti10}. It concludes that the only probabilistic score with these three natural properties is the logarithmic score, also often referred as the ignorance score (see~\cite{wilks19} and references therein). We will thus use the logarithmic score to validate the forecast skills of the neural network.

The aim of this section is to define the logarithmic score, to explain that it is nothing else than the negative of the cross-entropy loss function minimized by the neural network during the learning stage, and to define a normalized logarithmic score. The Normalized Logarithmic Score (NLS) is just a simple affine transformation of the logarithmic score that has the property of being equal to 0 when the forecasted probability is the climatological frequency, and equal to 1 when the prediction is perfect. 

We consider $N$ actually observed events $Y_{(n)}$, with $1\leq n \leq N$, associated with  the observed features $\X_n$. We suppose that the event $Y_{(n)}$ is observed pertaining to the class $k_n$. This means that $Y_{(n),k_n}  = 1$, and  $Y_{(n),l}  = 0$, for $l \neq k_n$. The couples $(\X_n,Y_{(n)})$ are identically distributed, and the probability that $Y_{(n);k_n}=1$ given that $\X_n=\x$ is $p_{k}(\x)=\mathbb{P}\left(Y_k=1\left|\X=\x\right.\right)$.

The real probabilities $p_{k}(\x)$ are unknown. We consider a probabilistic forecast $\hat{p}_k(\x)$. In our case, $\hat{p}_k(\x)$ will be the output of the  neural network after the learning stage. Our aim is to give a score that quantifies the quality of the approximation of $p$ by $\hat{p}$. This score should be computed without the actual knowledge of $p$, and be based only on the $N$ observations, for instance $N$ samples of a test dataset. The logarithmic score~\cite{benedetti10} is  
\begin{equation}
	S_N(\mathbf{\hat{p}}) =- \frac{1}{N}\sum_{n =1}^N \log(\hat{p}_{k_n}(\x_n))
\end{equation}
We note that in the simple case considered in~\cite{benedetti10}, the probabilities do not depend on the state of the system $\x$, while they do in this paper. However, all the reasoning and conclusions in~\cite{benedetti10} easily generalize to this new case. We note that with this sign convention, the minus sign in front of the logarithm, the logarithmic score is positive $(S_N >0)$ and negatively oriented (the smaller the score, the better the prediction).\\

We see that the logarithmic score is nothing more than the empirical cross-entropy loss function:
\begin{equation}
	{
		\mathcal{C} [\hat{\mathbf{p}}] =
		-\frac{1}{N}\sum_{n=1}^{N} \sum_{k=0}^{K-1} \delta_{k_n, k} \log \hat{p}_k\left( \boldsymbol{x}_{n} \right), 
		\label{eq:CrossEntropy}
	}
\end{equation}
where $\delta_{k_n, k}$ is the Kronecker delta. This is the loss function minimized during the learning stage of the neural network.

It is easy to check that the score is proper, by noting that according to the law of large numbers $\underset{N \rightarrow \infty}{\lim} S_N = \mathbb{E}[S_N] = L\left[ \hat{\mathbf{p}}, \mathbf{p} \right]$, with 
\begin{equation}
	L\left[ \hat{\mathbf{p}}, \mathbf{p} \right] = -\int {\rm d}\x\ P(\x)\sum_{k=0}^{K-1}p_{k}(\x)\log\left[\hat{p}_{k}(\x)\right],
	\label{eq:Loss function}
\end{equation} 
and that the minimum of $L\left[\hat{\mathbf{p}},\mathbf{p}\right]$ is obtained for $\hat{\mathbf{p}} = \mathbf{p}$. 

We note that $L\left[\mathbf{p},\mathbf{p}\right] \geq 0$ and that $L\left[\mathbf{p},\mathbf{p}\right] = 0$ only in the case when for any $\x$, all the $p_k(\x)$ are equal to zero except one. This is the case of a deterministic relation between $\x$ and $y$, when a perfect prediction is possible. In general, when the relation between $\x$ and $y$ is stochastic, $L\left[\mathbf{p},\mathbf{p}\right] > 0$, and it measures the level of stochasticity between  $\x$ and $y$.\\

It is important to compare the obtained score with the score of a prediction based on the climatological frequency. We define the climatological frequency as the probability $\overline{p}_k$ of observing the class $k$, independently of the knowledge of the state of the system $\x$. The climatological forecast $\hat{\mathbf{p}} = \overline{\mathbf{p}}$ provides a baseline: any skillful forecast should be better than the climatological one. We note that $\mathbb{E}[S_N(\overline{\mathbf{p}})]=L\left[ \overline{\mathbf{p}}, \overline{\mathbf{p}} \right] =- \sum_k \overline{p}_k \log \overline{p}_k$ (we have used (\ref{eq:Loss function}), noting that the $\overline{p}_k$s do not depend on $\x$). 

Generalizing the discussion in~\cite{benedetti10} to the present case, we define the  \emph{Normalized Logarithmic Score} as 
\begin{equation}\label{eq:NSS}
	\text{NLS}(\hat{\mathbf{p}}) = \frac{-\sum_k \overline{p}_k \log \overline{p}_k - S_N(\hat{\mathbf{p}})}{-\sum_k \overline{p}_k \log \overline{p}_k}.
\end{equation}
We clearly see that for the climatological forecast $\mathbb{E}[\text{NLS}(\overline{\mathbf{p}})]=0$. As $-\sum_k \overline{p}_k >0$, we see that the score is positively oriented (the larger score, the better prediction). 

For a given $p$, the optimal value of $\mathbb{E}\left[\text{NLS}(\hat{\mathbf{p}})\right]$ is $\left(-\sum_k \overline{p}_k \log \overline{p}_k - L\left[\mathbf{p},\mathbf{p}\right]\right)/\left(-\sum_k \overline{p}_k \log \overline{p}_k\right) \leq 1$. This optimal value is unknown, except if one would know $\mathbf{p}(\x)$ and $P(\x)$. Only in the case of a deterministic relation between $\x$ and $y$, when a perfect prediction is possible, this optimal value is equal to 1, otherwise it is strictly smaller than one.

The Normalized Logarithmic Score is a way to quantify the predictive skill of predictive power of the models learned by the convolutional neural networks. We conclude that the \emph{Normalized Logarithmic Score} is positively oriented (the larger the better), its average is equal to zero for the climatological forecast, and is always smaller than 1. The unknown optimal value is strictly smaller than 1, except when a deterministic relation between the predictor $\x$ and the predicted class $y$ exists and a perfect prediction is possible. These  properties make it convenient.\\

In the climate and meteorology literature, other scores for probabilistic forecasts are commonly used, for instance the Brier Score~\cite{Brier50,wilks19}. The Brier score can be very useful. It however depends on unobserved events (see~\cite{benedetti10}). Moreover the relation between the logarithmic score and the cross-entropy loss function makes the learning and the validation stage fully compatible. The logarithmic score also appears to be more sensitive towards the measurements of small probabilities or probabilities close to one, and is thus better suited for the study of rare events. The information theoretic interpretation of the logarithmic score is also an appealing property.



\section{Neural network architecture and learning protocol for the prediction of rare event probabilities}
\label{sec:NN}

\begin{figure*}
	\centering
	\tdplotsetmaincoords{60}{50}
	\begin{tikzpicture}[tdplot_main_coords,line join=miter,font=\sffamily]
		\tikzset{Relu/.style={red, draw=blue, fill=green!20, minimum width=1.5cm,minimum height=0.75cm}}
		\tikzset{ReluFlat/.style={red, draw=blue, fill=yellow!20, minimum width=1.5cm,minimum height=0.75cm}}
		\tikzset{Softmax/.style={red, draw=blue, ultra thick, fill=red!30, minimum width=2cm,minimum height=1cm}}
		\path[tdplot_screen_coords] (-2,0); 
		\pgfmathsetmacro{\xinitialstretch}{4.5} 
		\pgfmathsetmacro{\xstretchchange}{0.25} 
		\edef\Cols{red,red,dgreen,blue,purple,green,yellow} 
		\edef\LstCols{{"red","red","dgreen","blue","purple","green","yellow"}} 
		\pgfmathsetmacro{\yslope}{0.4} 
		\foreach \Col [count=\X starting from 0,remember=\X as \LastX] in \Cols 
		{
			\pgfmathsetmacro{\xstretch}{\xinitialstretch-\xstretchchange*\X} 
			\pgfmathsetmacro{\Xp}{2+2*\X} 
			{\foreach \XX in {0,1,...,\Xp} 
				{
					\pgfmathsetmacro{\pagesep}{0.28} 
					\pgfmathsetmacro{\pagesepchange}{0.05} 
					\pgfmathsetmacro{\pageloc}{\xstretch*\X+1*\pagesep*\XX-1*\pagesepchange*\XX*\X} 
					\ifnum\X=0 
					\pgfmathsetmacro{\pagelocz}{2-2*\XX} 
					\else 
					\pgfmathsetmacro{\pagelocz}{0}
					\fi
					\def\pagelocstart{\xstretch*\X+1*\pagesep*\XX-1*\pagesepchange*\XX*\X-1*\pagesep} 
					\begin{scope}[canvas is yz plane at x=\pageloc]
						\pgfmathtruncatemacro{\fullness}{120-20*\XX} 
						\ifnum\X>0
						\draw[fill=\Col!\fullness] (-1.8+0.3*\X,-1.8+0.3*\X+\pagelocz) rectangle (1.8-0.3*\X,1.8-0.3*\X+\pagelocz); 
						\else
						\ifnum\XX=0
						\node (BN-\X) at (-3,2) {$T_F$};
						\draw[fill overzoom image=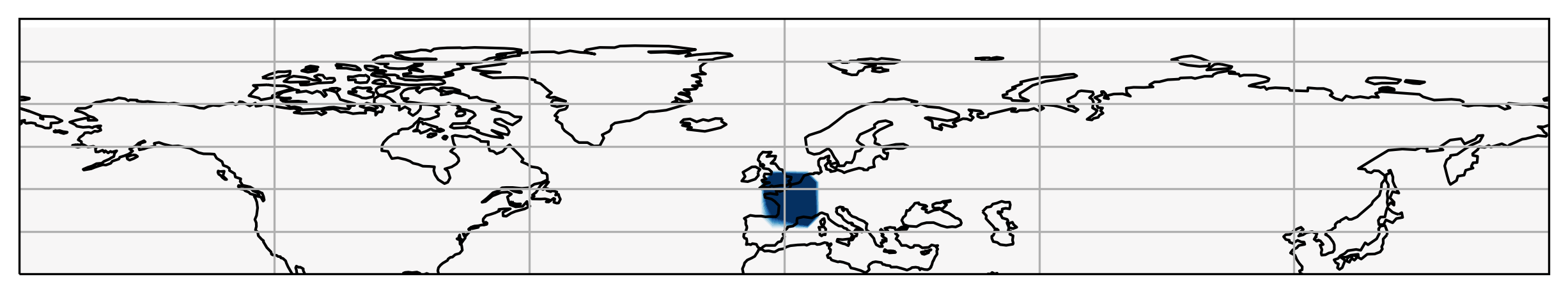] (-2.5,-.8+\pagelocz) rectangle (2.5,.8+\pagelocz); 
						\fi
						\ifnum\XX=1
						\node (BN-\X) at (-3,0) {$Z_{NH}$};
						\draw[fill overzoom image=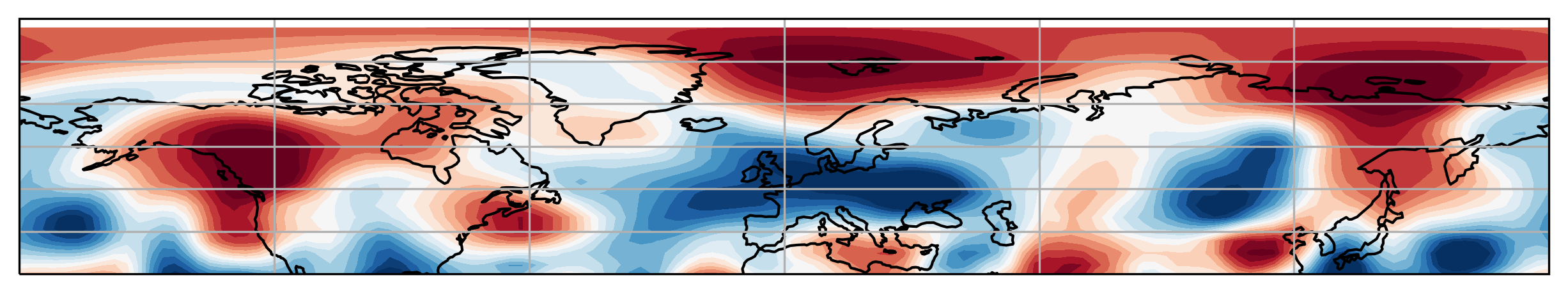] (-2.5,-.8+\pagelocz) rectangle (2.5,.8+\pagelocz); 
						\fi
						\ifnum\XX=2
						\node (BN-\X) at (-3,-2) {$S_F$};
						\draw[fill overzoom image=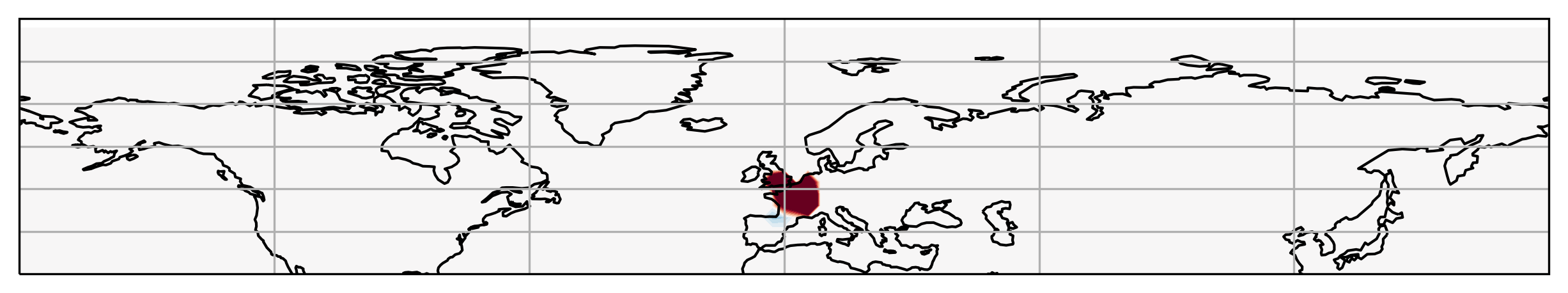] (-2.5,-.8+\pagelocz) rectangle (2.5,.8+\pagelocz); 
						\fi
						\fi
					\end{scope}
					\draw[thick,\Col] (\pagelocstart,0,0) -- (\pageloc,0,0); 
				}

				\ifnum\X=0
				\node[anchor=west, rotate=-20] at (\xstretch*\X,3-0.25*\X,1){Conv $2D_\X (3\times3\times 32)$}; 
				\fi
				\ifnum\X=1
				\node[anchor=west, rotate=-20] at (\xstretch*\X,3-0.25*\X,1){Max Pool $2D_\X (2\times2)$}; 
				\fi
				\ifnum\X=2
				\node[anchor=west, rotate=-20] at (\xstretch*\X,3-0.25*\X,1){Conv $2D_\X (3\times3\times 64)$}; 
				\fi
				\ifnum\X=3
				\node[anchor=west, rotate=-20] at (\xstretch*\X,3-0.25*\X,1){Max Pool $2D_\X (2\times2)$}; 
				\fi
				\ifnum\X=4
				\node[anchor=west, rotate=-20] at (\xstretch*\X,3-0.25*\X,1){Conv $2D_\X (3\times3\times 64)$}; 
				\fi
				
				\draw[thick,\Col] (\xstretch*\X+0.4*4,0,0) coordinate(front-\X) 
				-- (\xstretch*\X+2+0.4*4,0,0) coordinate(back-\X); 
				\begin{scope}[canvas is yz plane at x=\xstretch*\X+0.4*4,transform shape] 
					\ifnum\X=6
					\node[Softmax] (BN-\X) at (0,0) {Softmax};
					\else
					\ifnum\X=0
					\node[Relu] (BN-\X) at (0,0) {ReLu};
					\fi
					\ifnum\X=2
					\node[Relu] (BN-\X) at (0,0) {ReLu};
					\fi
					\ifnum\X=4
					\node[ReluFlat] (BN-\X) at (0,0) {ReLu+Flat};
					\fi
					\ifnum\X=5
					\node[ReluFlat] (BN-\X) at (-.75,0) {N1};
					\node[ReluFlat] (BN-\X) at (.75,0) {N2};
					\fi
					\fi
				\end{scope}
				\pgfmathtruncatemacro{\NextX}{\X+1} 
				\ifnum\X<4
				\pgfmathsetmacro{\NextCol}{\LstCols[\NextX]} 
				\else 
				\pgfmathsetmacro{\NextCol}{"black"} 
				\fi 
				\draw[very thick,\NextCol,->>] (\xstretch*\X+0.4*4,0,0) -- ({\xstretch*(\X+1)+0.4*4},0,0); 
			}
		}
	\end{tikzpicture}
	\caption{Schematics of the CNN architecture. The numbers {$a\times b\times c$} inside the boxes for Conv2D indicate the kernel sizes (a by b) and the number of filters $c$ in the convolutional layers. The box ``Relu'' actually consists of a sequence: batch normalization - activation (ReLu), and spatial dropout. Relu stands for Rectified Linear unit, it provides non-linearity and is typically used to prevent vanishing gradients instead of sigmoid for the hidden layers. 
	}
	\label{fig:architecture}
\end{figure*}
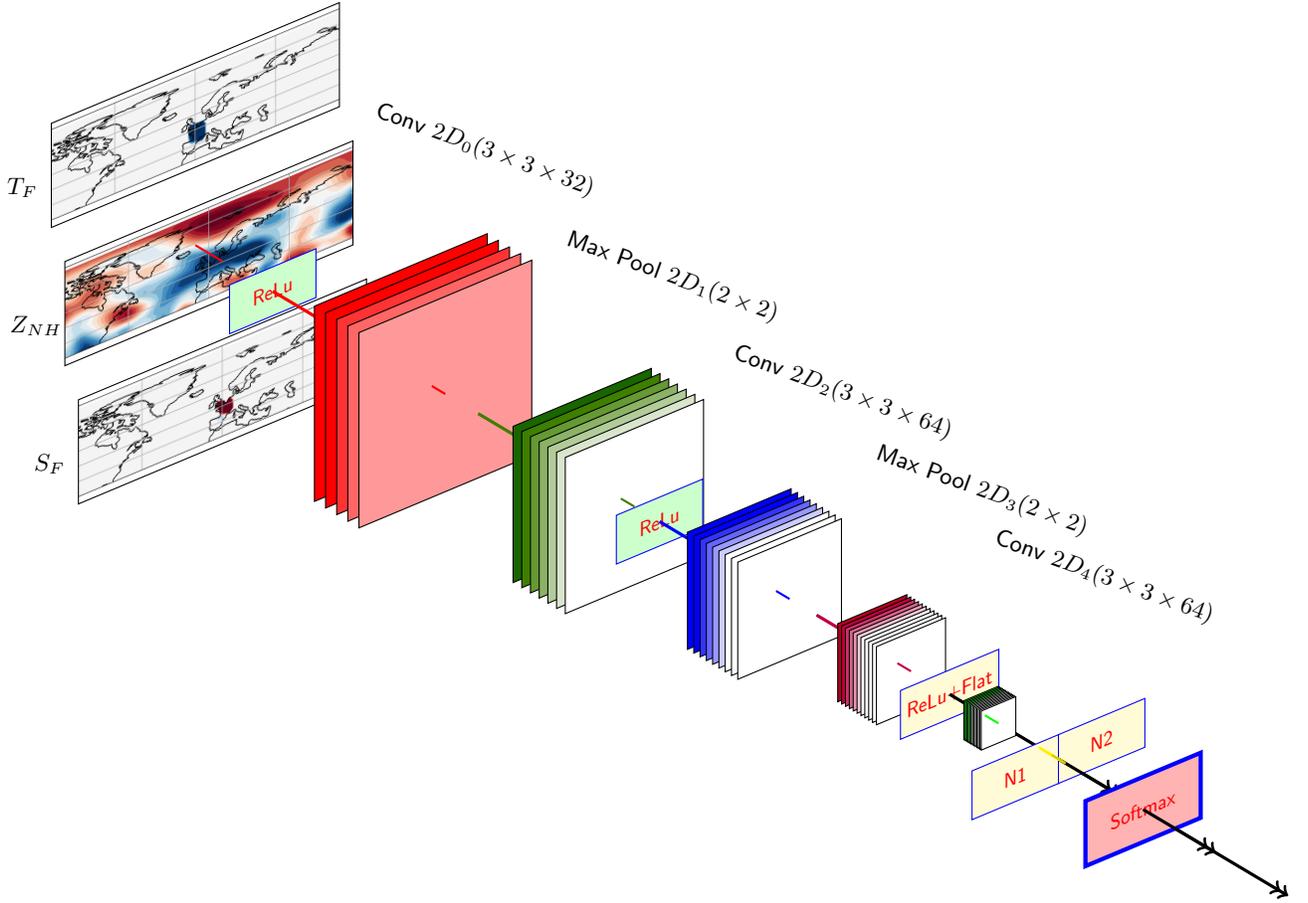

As explained in Section 2, the task is to predict the probability of occurence of a heatwave at time $t$, from the state of the system  $\x$. 
These variables are sets of physical fields which are observed at time $t-\tau$. 
In this section, we present the neural network architecture, the training parameters and protocols.

\subsection{Neural network Architecture and learning parameters}
\label{NNarchitecture}

\noindent \textit{Neural network Architecture.} 
For the inference of probabilities, we will use a Convolutional Neural Network (CNN) architecture as sketched in Figure~\ref{fig:architecture}~\cite{Goodfellow-et-al-2016}.

It consists of a 3-layer architecture, combining convolutional filters, followed by ReLu activations. 
For all layers, $3 \times 3$ convolution kernels are used, while the number of filters is $32$ for the first and $64$ for the two last {ones}. 
Two max-pool operations are inserted between the convolution layers after the activation functions. The output of the third layer is flattened and used as input of a dense layer with 64 neurons, a layer with 2 outputs corresponding to the heatwave labels, and a softmax function which maps the outputs to (0,1) range, as detailed for the probabilistic interpretation in Section~\ref{sec:predprob}. 

The probabilities obtained  by softmax regression with cross-entropy loss function are not always well calibrated. For instance, it has been discussed in~\cite{guo17} in very deep networks (e.g., ResNET) that the calibration of the probabilities is not correct; in other words the network may be overly confident about its probabilistic predictive capability.  This is a reason why we prefer to use a neural network that is not too deep.  Also, this phenomenon can be worse when  facing extremely rare events because of the imbalance between the classes.
We discuss in Section~\ref{ssec:EarlyStopping} how to avoid biases due to overfitting.

\noindent \textit{Loss function.} 
As discussed in section~\ref{sec:predprob}, we minimize the cross-entropy loss function, Eq.~(\ref{eq:CrossEntropy}). Optimizing the cross-entropy is done as a supervised task, using both the input data fields $\x$ and  heatwave classes ${Y}$. 

\noindent \textit{Learning tools and parameters.} 
The CNN layers are implemented using Tensorflow 2 package, and CNN training is done with Adam optimizer,  with learning rate set to $2 \cdot 10^{-4}$.
Network weights are initialized using a standard Glorot distribution. The computer resources consisted of  computers with dedicated graphics cards such as GV100GL [Tesla V100 PCIe 16GB], TU102 [RTX 2080 Ti Rev. A] and TU104 [RTX 2080 SUPER].

\subsection{Training protocol} 
\label{sec:training}  \label{ssec:training}

\noindent {\bf Data normalization.} 
The training set consists of 8,000 independent and statistically equivalent years of simulated climate (cf. section~\ref{sec:plasimdata}). 
The season of interest is June-July-August (JJA, 90 days). When we consider 14-day time average, we have 77 days of JJA per year before the start of the heatwave (see section~\ref{sec:defheatwave}). This gives 616000 snapshots for each value of $\tau$. 

Data are normalized by grid point prior to application to the neural network:
we add a constant and scale each cell of each field such that the sample mean and variance are $0$ and $1$, respectively.

Note that each possible input field, $Z$, $S$ or $T$ is considered as a separate input channel and they are stacked in the CNN layers (e.g., like RGB channels in colors images). The input tensor is then of size $22\times128\times$ the number of input fields (3 if using all of $Z$, $S$, and $T$) provided that the global field corresponding to the north hemisphere above 30N is used. In this case we use the notation $Z_{NH}, T_{NH}, S_{NH}$. On the other hand if the input consists of a smaller area corresponding to the North Atlantic region and Europe we use the notation $Z_{NAE}, T_{NAE}, S_{NAE}$ which has dimensions of $18 \times 42$ (Fig~\ref{fig:NorthHemisphere}). Note that in both cases we could be interested in applying additional mask to the area of France, i.e. setting to zero all values external to the area of a box around France. This operation is not applied to geopotential, so the two resulting cases are  $Z_{NH}, T_{F}, {S_F}$ and  $Z_{NAE}, T_{F}, {S_F}$.

Sec.~\ref{sec:results} will discuss in details which combination of these  $z$, $s$ and/or $T$ fields, used either locally or globally give the best predictions. 

\noindent {\bf Stratified 10-fold cross-validation.}
To quantify confidence in achieved prediction performance, we  use a classical stratified k-fold cross-validation procedure~\cite{hastie2009elements}. 
The 8000 available years are randomly split into $k=10$ subsets.
The splitting of the data set is performed on a per-year basis to avoid that any of the years is split into a test and train set. The latter would blur validation consistency by spurious temporal correlations between validation and train sets or seasonal effects.
In addition, random splitting is performed so as to maintain the same number of heatwaves per subset (stratification). Each resulting subset consists of 800 years. 

\noindent {\bf Initialization, batch normalization and dropout.} 
The convolutional and ReLu activation layers are followed by batch normalization and subjected to a dropout.
Batch normalization is expected to accelerate learning.
Dropout  is considered a regularization tool avoiding overfitting~\cite{guo17}. 
The dropout rate is set to $0.25$.

\subsection{Early Stopping}
\label{ssec:EarlyStopping}

Overfitting is a major problem of machine learning, especially in the case of deep neural networks. This means that the model is trained to reproduce the training dataset too closely and does not generalize well to the validation set or the test set. As a remedy an early-stopping strategy is typically used, which implies stopping the training at an epoch when the appropriate metric on the validation set starts getting worse. We have followed the same general outline, with a caveat. 
Since we are performing 10-fold cross validation, we can rely on the performance metric as provided by the Normalized Logarithmic Score (NLS), equation ~\eqref{eq:NSS}. Each training set (fold) reaches its optimal performance on a validation set at a certain epoch, after which it starts to over-fit. 
This epoch generally depends on the training set although in most cases the variance is not vary large. 

Nevertheless, we propose to select an epoch on which the fold-wise mean score is optimal. We refer to this as collective early stopping.
This approach is more conservative about the performance of the network, and less dependent on the training/validation split.

\subsection{Unbalanced classes and undersampling strategy for probabilistic forecast}\label{sec:unbalancedunbiasing}

Rare event prediction, suffers by design from a severe class imbalance. 
To address this, we use a majority 
class undersampling strategy, as opposed to a minority 
class oversampling. This means that the training set uses all available positive events, but only a ratio $1/r$ of events of the majority events are drawn with uniform probability, with $r>1$ (undersampling). This {data reduction} procedure 
also minimizes time and memory costs during training. 

This undersampling changes the rate of positive events. We need to take account of this change of measure, otherwise the predicted probability will not be correct \cite{fernandez2018learning,pozzolo15a}.

Let $p_{0}(x)$ and $p_{1}(x)=1-p_{0}(x)$ denote the probabilities that $Y=0$ and $Y=1$, respectively, given that $X=x$, in the original set. Let $p_{0}^{\prime}(x)$ and $p_{1}^{\prime}(x)=1-p_{0}^{\prime}(x)$ denote the probabilities, that $Y=0$ and $Y=1$, respectively, given $X=x$, in the undersampled training set. These probabilities are obviously related as: 
\begin{equation}
p_{0}^{\prime}(x)=\frac{p_{0}(x)}{p_{0}(x)+r(1-p_{0}(x))}\text{\,\,\,and}\,\,\,p_{1}^{\prime}(x)=\frac{rp_{1}(x)}{1-p_{1}(x)+rp_{1}(x)}.\label{eq:p_pT}
\end{equation}
As an example, when $p_{0}=0.8$ and $p_{1}=0.2$ and an undersampling
ratio $r=4$ is used, fully balanced undersampled classes are obtained, with
$p_{0}^{\prime}=0.5$ and $p_{1}^{\prime}=0.5$. 
During training after undersampling, the  neural network actually gives an estimate $\hat{p}_{0}^{\prime}(\x)$ and $\hat{p}_{1}^{\prime}(\x)$ of the probabilities $p_{0}^{\prime}(\x)$ and $p_{1}^{\prime}(\x)$ of the event that has been seen, and not of the true ones $p_{0}(\x)$ and $p_{1}(\x)$. 

In order to get an estimate $\hat{p}_{0}(\x)$ and $\hat{p}_{1}(\x)$ of  $p_{0}(\x)$ and $p_{1}(\x)$, respectively, we need to invert the relation~\eqref{eq:p_pT} between the initial probabilities and the probabilities in the undersampled set. This gives
\begin{equation}
\hat{p}_{0}(x)=\frac{r\hat{p}_{0}^{\prime}(x)}{1-(1-r)\hat{p}_{0}^{\prime}(x)}\text{\,\,\,and}\,\,\,\hat{p}_{1}(x)=\frac{\hat{p}_{1}^{\prime}(x)}{r+(1-r)\hat{p}_{1}^{\prime}(x)}.
\label{eq:pT_p}
\end{equation}

The estimated probabilities $p_{0}^{\prime}(\x)$ and $p_{1}^{\prime}(\x)$ can then be tested using~\eqref{eq:NSS} on a validation set or a test set, or used as the predicted probabilities for the physical discussion and for applications. Please note that we do not undersample the validation set.

In this work, we use an undersampling rate $r=10$,  consistent with~\cite{jacques-dumas22}. This reduces the RAM memory usage approximately 10 fold and accelerates the training while not impacting the skill significantly (see Figure~\ref{fig:Undersampling_Skill} in appendix~\ref{app:undersampling}).



\section{Probabilistic forecast of extreme heatwaves}
\label{sec:results}

The present section aims to quantify, using the Normalized Logarithmic Score $NLS$, equation~(\ref{eq:NSS})), the quality of the prediction of heatwave occurence probabilities. 
These quantifications will be conducted as function of the lag $\tau$ between time at which data is available for prediction and heatwave occurrence. 
The impact of the nature of the data used for prediction (soil moisture, geopotential and/or 2-m temperature) will be investigated together with the benefits of possible combinations of such inputs.
Further, the impact of the amount of available  data on prediction performed will be studied. 

\subsection{Relevant climate fields for the probabilistic forecast of extreme heatwaves}
\label{field_comparison}

A first key question is to assess which of the predictors, among the physical and dynamical fields, have the best prediction capabilities. To this aim, we train the neural network with the large 8,000-year dataset and test its skill by computing the Normalized Logarithmic Score  (cf. equation~\eqref{eq:NSS}) on a validation set. From the values of $NLS$, we can compare the prediction skills for extreme heatwaves, when different combinations of the fields are used: soil moisture $S$, geopotential height at 500-hPa $Z$, and 2-m temperature $T_{2m}$, used either alone, or combined.

In this section we always use the geopotential height at 500-hPa $Z_{NH}$ over the Northern Hemisphere area, and the soil moisture $S_F$ and 2-meter temperature $T_F$ over the France area. These choices of regional masking will prove to be the optimal ones, as we will discuss in section \ref{sec:datasize}. We also consider the 2-meter temperature integrated over the France area, which is then a single real number denoted $T_{FI}$. When only one local field ({FI}) is used, a simple scalar logistic regression is performed. For all the other cases, we train the neural network as explained in section~\ref{sec:inputs}.

\begin{figure*}[h]
	\centering
	\subfloat[\label{fig:univariate}]{%
		\centering
		\includegraphics[width=0.3\textwidth]{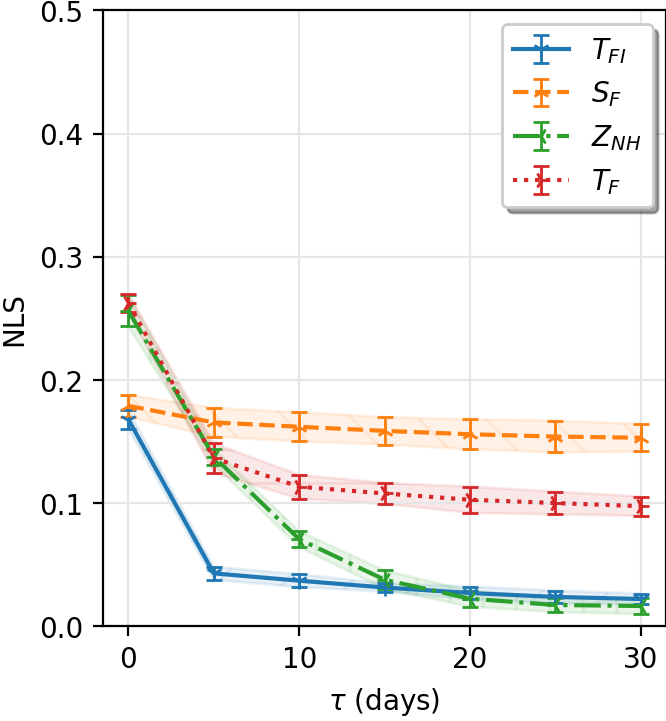}
	}\hfill
	\subfloat[\label{fig:pairwise}]{%
		\centering
		\includegraphics[width=0.3\textwidth]{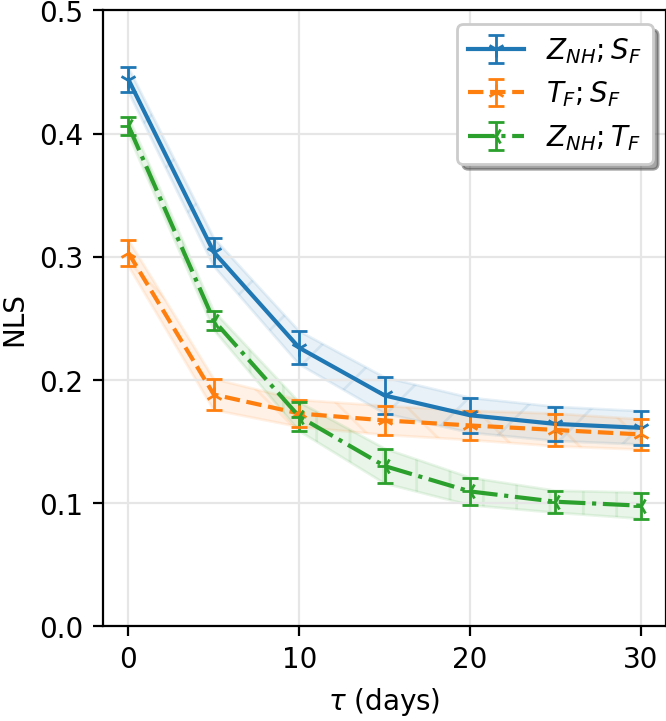}
	}\hfill
	\subfloat[\label{fig:3wise}]{%
		\centering
		\includegraphics[width=0.3\textwidth]{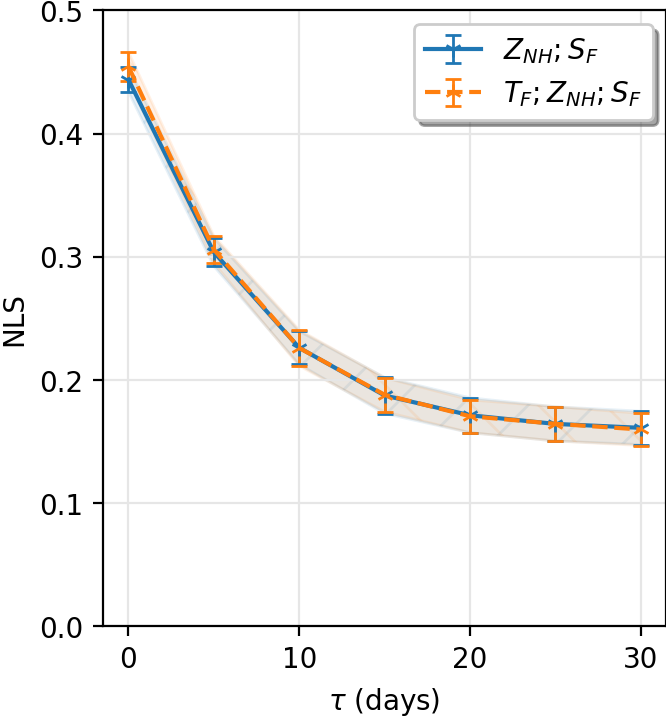}
	}
	\caption{{\bf Relevant climate fields for best prediction:} All figures show the Normalized Logarithmic Score $NLS$ (cf. equation~\eqref{eq:NSS}), versus the lead time $\tau$, for different combinations of the predictor fields. Fields are either integrated over the area of France (FI), masked over the France area (F) or over the Northern Hemisphere (NH). From $k-$fold cross-validation values, we plot the averaged $NLS$, plus or minus one standard deviation (shaded area and error bars): (a) Single field prediction: ${T_{FI}}$ (blue), ${S_F}$ (orange), ${Z_{NH}}$ (green), and ${T_F}$ (red)~; (b) Fields combined pairwise for prediction: ${(Z_{NH}, S_F)}$ (blue), ${(T_F, S_F)}$ (orange), ${(Z_{NH}, T_F)}$ (green),~; (c)  All three fields combined for prediction: ${(T_F, Z_{NH}, S_F)}$ (orange).} 
	\label{fig:DifferentFieldsComparisons}
\end{figure*}

Figure~\ref{fig:DifferentFieldsComparisons} reports the Normalized Logarithmic Scores versus the lead time $\tau$, for different combinations of fields.  

\noindent \textit{Single Field Prediction.} 
Figure~\ref{fig:univariate} first shows that soil moisture over France, $S_F$, conveys significant long-term ability for the prediction of heatwave probabilities. In particular, it retains predictive value at large lead time $\tau$, hence can be considered a candidate slow physical driver. Those findings are consistent with the important role of soil moisture through its two-way coupling with heatwaves as discussed in section \ref{sec:inputs}. The main physical interpretation is that soil moisture deficit statistically increases the temperature in the lower troposphere, through the impact of deficit of evopotranspiration or other water exchanges at the surface on the lower troposphere energy cycle, and on cloud cover. We also note that the predictive skill of soil moisture alone only weakly decays with lead time $\tau$, which can be interpreted as a consequence of the long correlation time of soil moisture compared to the maximum 30 days lag-times considered here, soil moisture being a stock. The soil moisture predictability skill is nearly a constant versus $\tau$. This constant could be directly related to the conditional probability to have a heatwave given some soil moisture field $S_F$, and could be estimated in a straightforward fashion based on a climatological prediction conditioned on the values of the soil moisture fields, regardless of any other information about dynamics.  

The curve $Z_{NH}$ on Figure~\ref{fig:univariate}, shows that when using only the Northern Hemisphere 500 hPa geopotential height for training, the neural network has better prediction skill for short times. This skills decays roughly exponentially with $\tau$ 
with the approximate rate of decay of $0.13$ per day: $NLS_{z_G}\approx 0.26 \exp(-0.13\, \tau)$. This rate corresponds to a decay time of 7.7 days (after 7.7 days, the prediction skill decreased by a factor $e$). The 500 hPa geopotential height field is considered as one of physical fields which characterizes the best midlatitude troposphere dynamics, Rossby waves, cyclonic and anticyclonic anomalies. The decay of the skill is interpreted as the progressive loss of memory for the evolution of this dynamical field
, due to the chaotic dynamics of the midlatitude troposphere, over time scales of the order of the synoptic time scale, which corresponds to few days. 

By comparing the Normalized Logarithmic Score $NLS$ from $Z_{NH}$ and $S_F$, we confirm that in the short run soil moisture has significantly smaller predictive value than the 500 hPa geopotential height, but, in contrast, it keeps its predictive skill over much longer time lags $\tau$, as expected based on the discussion in section \ref{sec:inputs}. However, such qualitative statements are now precisely quantified, thanks to the neural network we introduced here. We see for instance that, for these PlaSim simulations, the geopotential height $Z_{NH}$ alone has a $NLS$ of $0.26\pm0.01$ compared to $0.18\pm0.01$ for soil moisture $S_F$ at $\tau=0$, and that the predictive skill of soil moisture alone becomes larger than the one of 500 hPa geopotential height for $\tau$ larger than about 4 days.

The interest of looking at the predictive power of the temperature integrated over France, $T_{FI}$, is to assess the predictability properties related to persistence and possibly low tropospheric advection. $T_{FI}$ on Figure~\ref{fig:univariate} shows this  prediction for $\tau=0$. Although not visible on the Figure, the related skill extends for $\tau$ of order of a few days, subsequently the predictability power is lost. For larger values of $\tau$, $\tau \geq 5$ days,  $T_{FI}$ reaches a weakly decreasing plateau. 

The information contained in $T_{FI}$ is the spatial average of the temperature field over France $T_F$. As a consequence, $T_F$ on Figure~\ref{fig:univariate} has a higher NLS. The fact that the predictive skill for $T_F$ is much better than the one for $T_{FI}$ shows that the details of the spatial pattern of temperatures over France matter much for the prediction of extreme heatwaves defined globally over France. This is a very interesting result. Anticipating the following discussion, we will interpret the predictive skills of both $T_F$ and $T_{FI}$, for $\tau \geq 5$, to be due to their mutual information with the soil moisture. We can then interpret the better predictive skill of the spatial field $T_F$, compared to the one of $T_{FI}$, as the consequence of the larger information content of the soil moisture on some specific areas. This interpretation is consistent with past studies that argued that soil moisture matters more in areas prone to its deficits, rather than areas where soil is unlikely to dry. This interpretation should be studied further in future works. 

The two plateaus with strictly positive skills for both $T_F$ and $T_{FI}$ for large time lag $\tau$ are striking. We cannot expect those skills, that extend over time scales much longer than the synoptic times, to be related to persistent properties or to free troposphere dynamics, because they extend over time scales much longer than the mixing time of the uncoupled troposphere dynamics. This long-term skills might be related to some correlations between $T_{FI}$ and some slow physical drivers. We might hypothesize that $T_F$ and $T_{FI}$ contain statistical information related to the soil moisture. In order to study this hypothesis, we will now study the predictive skills of combined fields.\\

\noindent \textit{Predictions using fields combined pairwise.} 
	We now study the predictability skills of the  neural network when trained using combinations of the two fields. The results, reported on Figure~\ref{fig:pairwise}, show that the best combination is the couple $(Z_{NH},S_F)$. Compared to the results on Figure~\ref{fig:univariate}, it is striking to see that the predictive skills of $Z_{NH}$ and $S_F$ seem to add up~\footnote{``Add up'' is used here qualitatively, there is no mathematical reason why skills should actually add up arithmetically}. The curve can be approximated as $0.288*\exp(-0.144 \, \tau) + 0.155$, with a decay time of about 6.9 days relatively close to the one obtained for the geopotential height alone. Using the couple $(Z_{NH},S_F)$, the neural network is able to conveniently retain the useful information for prediction, from both the fast dynamical field $Z_{NH}$ and the slow physical driver $S_F$ in a seamless way. 
	
	The predictive skill of the couple $(T_F,S_F)$ is the worst among the three couples for small lead times $\tau$, and is not better than the skill of the field $S_F$ alone for large lead times. For large lead times, this means that all the useful predictive information lies in $S_F$. This remark supports the idea that the plateau for $T_F$ in Figure ~\ref{fig:univariate} has to be interpreted as the predictive skill for the 2-meter temperature, as a consequence of its mutual information with the soil moisture. Moreover, for large lead times, clearly, the flow of information is from the soil moisture to the 2-meter temperature, as combining both fields do not give improvements with respect to soil moisture alone. However, the 2-meter temperature actually provides new complementary information for small lead times, most probably because of the skill associated with persistence or low-tropospheric advection. 
	
	It is interesting to note that the couple $(Z_{NH},T_F)$ performs rather well, and that the information of $Z_{NH}$ and $T_F$ seems to add up too, just like the one for $Z_{NH}$ and $S_F$ do. This is particularly striking for small lead times $\tau$. Indeed, one might have expected that the information about the 2-meter temperature might have been contained in the dynamical field $Z_{NH}$, for short lead times. This is however not the case, the better skill when combining the two fields clearly proves that the temperature field value contains relevant predictive information that was not included in the dynamical field $Z_{NH}$, even for small lead times. Where does this information come from? Is it related to slow physical drivers? In order to answer this question, we will need to compare the result for $(Z_{NH},S_F)$ with the neural network skill when all three fields are combined together for the training.\\ 
	
	\noindent \textit{Predictions combining all three fields together.} 
			Figure~\ref{fig:3wise} displays the prediction performance when using the three fields $(T_F,Z_{NH},S_F)$ together, and compares it to the results obtained with the best pair $(Z_{NH},S_F)$.
			There is no improvement for the predictive skill when adding the 2-meter temperature field $T_F$ to the geopotential height and soil moisture fields $(Z_{NH},S_F)$, except maybe for very short lead time $\tau <5$. The improvement for very short lead time is not statistically significant as it is within error bars. If this was not the case, we could interpret it as the effect of properties of persistence or of low-tropospheric advection of the 2-meter temperature field over France. This lack of improvement means that all useful information for prediction in the 2-meter field, is actually already contained in the $(Z_{NH},S_F)$ fields.\\
			
			Our first conclusion is that the best prediction is obtained when the  neural network is trained using the combined information from the Northern Hemisphere 500 hPa geopotential height field and the soil moisture over the France area. The  neural network is able to seamlessly combine the information of the fast dynamical driver, the 500 hPa geopotential height field, and the slow physical one, the soil moisture. The temperature field over France does not seem to convey complementary information to these two fields, except perhaps at  $\tau=0$. But even for $\tau=0$ the improvement is not statistically significant given the dataset: $0.455\pm 0.012$ vs $0.445\pm 0.010$.
			
			It is customary in other prediction studies for extreme heatwaves to use the local 2-meter temperature field. Given what we have observed this makes sense when the information about soil moisture is not available.
			\\
			


				\noindent \textit{Is it useful to consider more predictor fields?} 
				
				\begin{figure*}
					\centering
					\subfloat[\label{fig:Adding_time_steps}]{%
						\centering
						\includegraphics[width=8.6 cm]{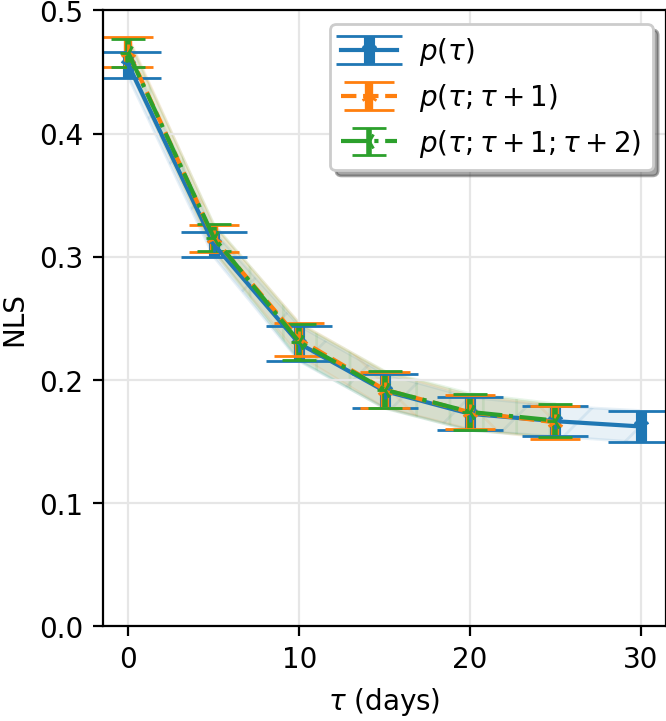}
					}\hfill
					\subfloat[\label{fig:Adding_zg}]{%
						\centering
						\includegraphics[width=8.6 cm]{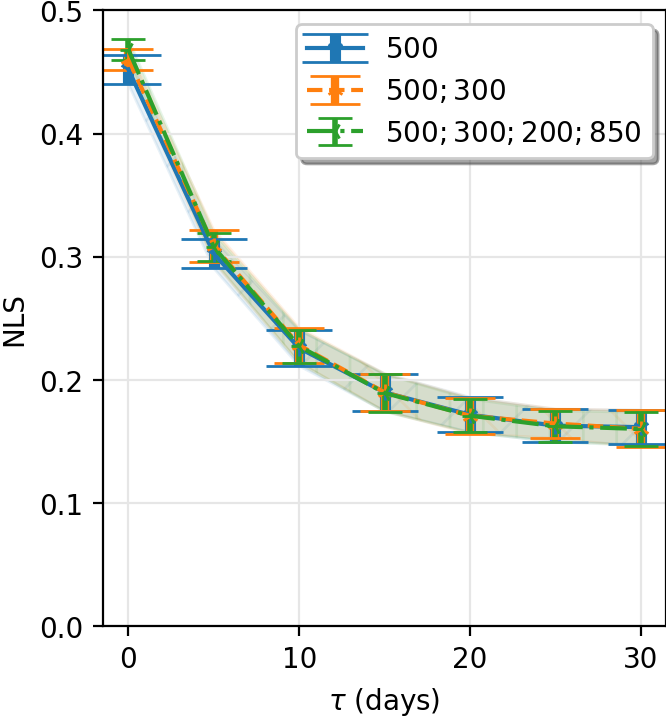}
					}
					\caption{Prediction skills when training the network with extra fields. The Normalized Logarithmic Score $NLS$ is on the $y$-axis. Panel (a) Addition of stacked fields at extra time steps: {$(T_F,Z_{NH},S_F)(t-\tau)$} (blue), {$\left[(T_F,Z_{NH},S_F)(t-\tau),(T_F,Z_{NH},S_F)(t-\tau-1)\right]$} (orange),{$\left[(T_F,Z_{NH},S_F)(t-\tau),(T_F,Z_{NH},S_F)(t-\tau-1),(T_F,Z_{NH},S_F)(t-\tau-2)\right]$} (green). The $NLS$ features a very small improvement when adding the previous day information, but none if we add two previous days. (b) Addition of extra levels of geopotential height {$\left[T_F,Z_{NH}(500 hPa),S_F\right]$} (blue), {$\left[T_F,Z_{NH}(500 hPa),Z_{NH}(300 hPa),S_F\right]$} {$\left[T_F,Z_{NH}(850 mbar),Z_{NH}(500 hPa),Z_{NH}(300 hPa),Z_{NH}(200 hPa),S_F\right]$} (green). The $NLS$ features a very small improvement, although not statistically significant, when adding more geopotential height fields.} 
					\label{fig:extrainput}
				\end{figure*}
				
				We now ask wether it might be useful to consider more predictor fields. We will train the neural network, first using the same fields but observed at more than one timestep, and second  considering the value of the geopotential height on other pressure isosurfaces. The two sets of results are visible on Figure~\ref{fig:extrainput}.
				
				We first train the neural network with the optimal set of fields $(T_F,Z_{NH},S_F)$, as in Figure~\ref{fig:3wise}. But during the training stage, rather than using the field values only at lead time $\tau$ (at time $t-\tau$) we also use the field values at lead time $\tau+1$ (previous day, at time $t-\tau-1$) and $\tau+2$ (second previous day). Those previous day fields are stacked with the fields at lead time $\tau$. Figure~\ref{fig:Adding_time_steps} shows the skill of the trained network adding the previous day fields (in orange), or the two previous day fields (in green). Adding the fields at previous timesteps is similar to delay-embedding in dynamical system theory~\cite{takens1981detecting}: the information lost in taking only part of the initial conditions of the deterministic dynamics can be recovered using fields at previous timesteps, in principle. The question we address here is more a practical one: can a given  neural network learn this missing information from the fields at previous timesteps, given the dataset length and its other practical limitations.
				
				The result in Figure~\ref{fig:Adding_time_steps}  shows a small statistically insignificant improvement when one adds the field values at lead time $\tau+1$, but no further improvement when one adds both the field values at lead time $\tau+1$ and $\tau+2$. This is a very interesting result. One can interpret this incapacity of given neural network to use the information at previous lead times in three different ways. The first possible interpretation, intrinsic to heatwave dynamics, would be that the gain in information content in the fields at previous lead times, to predict extreme heatwaves, is actually very small and within the error bars of our experiments. The second possible interpretation, practical in nature, would be that we have not found a network structure that could reliably recover this information. The third interpretation, would be that the 8,000 long dataset is too small in order for the neural network to practically learn such detailed information. Although we can not support precisely this claim with the present dataset, the analysis in the next section, of a lack of data regime, makes the third interpretation plausible.

				Rather than complementing the predictor fields with the ones at previous lead times, we now add other relevant dynamical fields at the same lead time $\tau$. Atmospheric and climate scientists know that the 500-hPa geopotential height field is rather relevant for dynamics, but that geopotential height at other altitudes or pressure isosurfaces are also useful and provide complementary information, for many phenomena. We now train the neural network with several sets of these fields, in addition to the optimal set of fields $(T_F,Z_{NH},S_F)$. The obtained skills are shown in Figure~\ref{fig:extrainput}{\color{blue}b}.
				
				The conclusion is that adding the geopotential height at 300 hPa (upper troposphere), orange curve, slightly improves the network prediction skill compared to the reference blue curve. However, this improvement is marginal, visible only for $\tau=0$, and even for $\tau=0$ it is within the error bar and thus not statistically significant. Similarly, we observe a minute increase in the Normalized Logarithmic Score when adding further the geopotential height at 850 hPa (lower troposphere), green curve. As for the case of delay embedding, the incapacity to improve the neural network prediction by adding more fields can be interpreted as being either intrinsic, or due to improper network architecture, or due to a lack of data for training. We suppose that the lack of data is the most plausible explanation. 
				
				Our second conclusion is that the set $(T_F,Z_{NH},S_F)$ or $(Z_{NH},S_F)$ are the optimal ones, with marginal difference in their predictive skills, for a dataset length of 8,000 years. The addition of any other extra fields including different lag-times prove to be superfluous.

			

			\subsection{Convergence of prediction skills with training dataset length and optimal areas for predictors: a regime of lack of data}
			\label{sec:datasize}

			\begin{figure}
				\centering
				\subfloat[\label{fig:NAE}]{%
					\centering
					\includegraphics[width=0.3\textwidth]{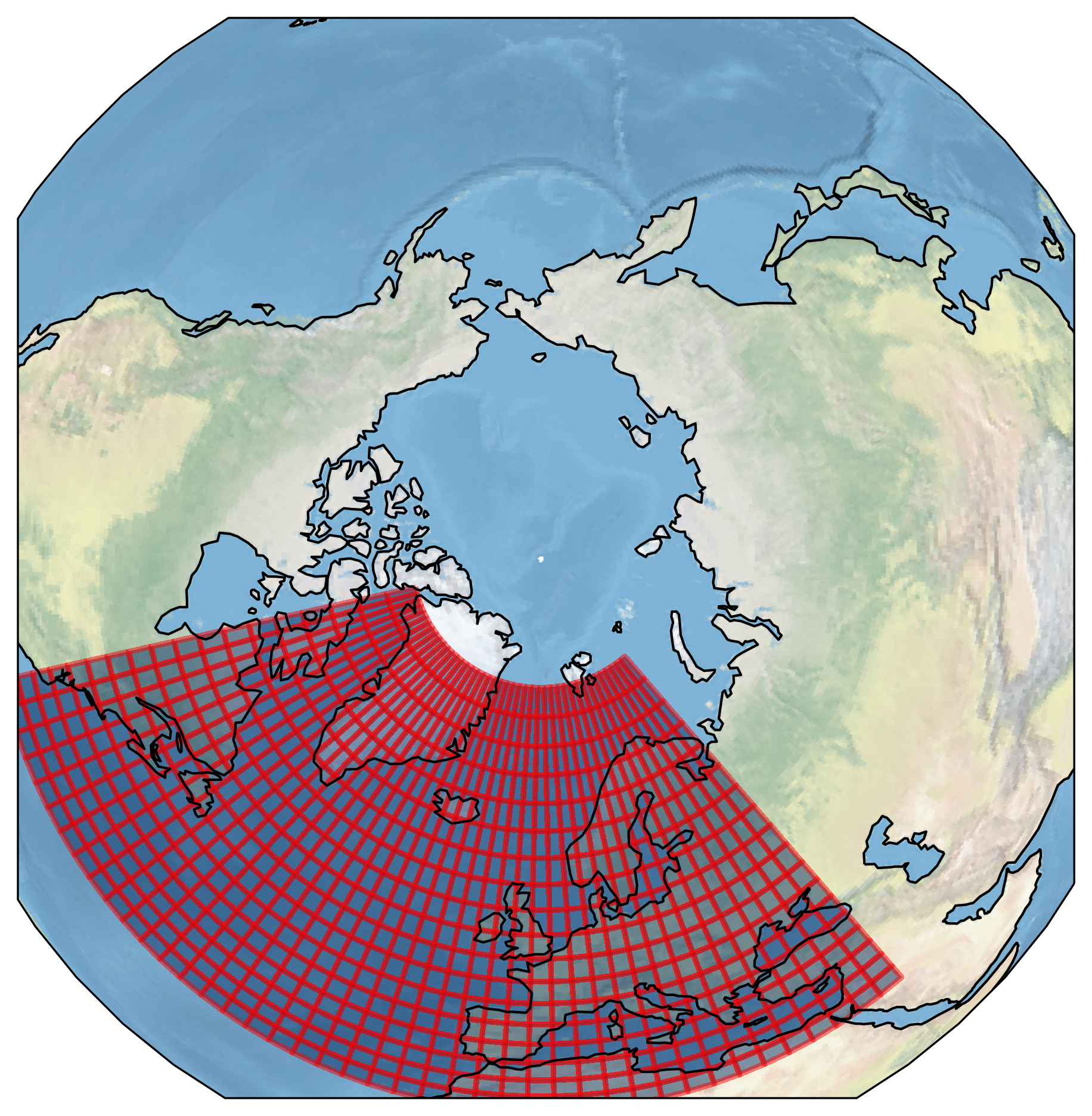}
				}\hfill
				\subfloat[\label{fig:datasize-b}]{%
					\centering
					\includegraphics[width=0.3\textwidth]{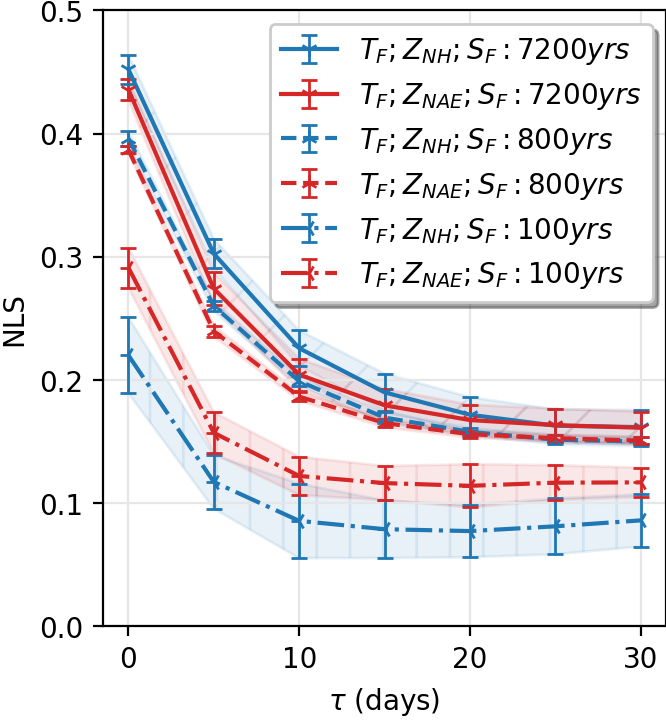}
				}\hfill
				\subfloat[\label{fig:NH}]{%
					\centering
					\includegraphics[width=0.3\textwidth]{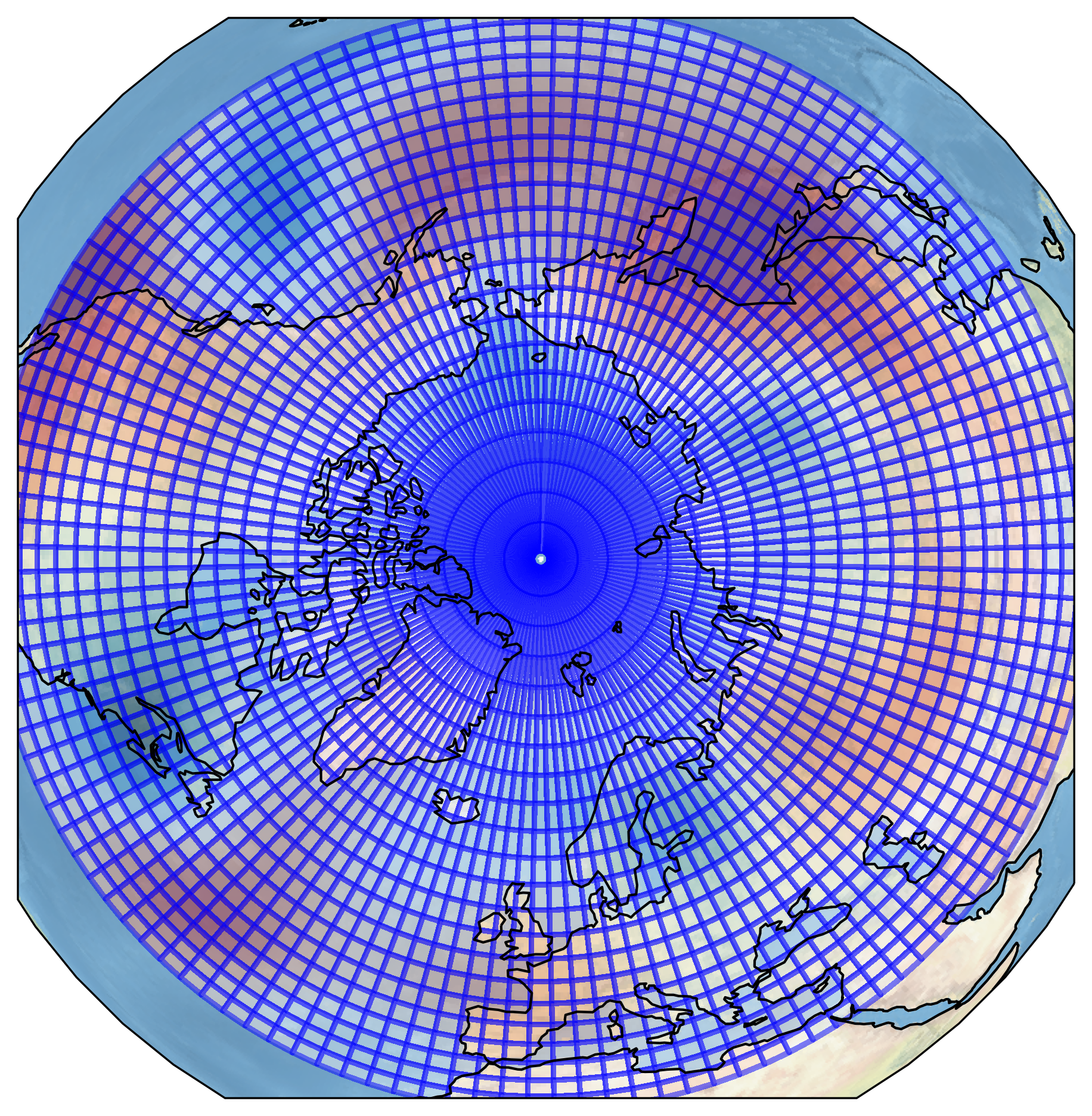}
				}
				\caption{{Prediction skills versus dataset lengths and optimal geographical area. Panels a) and c) show PlaSim grid for the North Atlantic and Europe sector (NAE,a), and Northern Hemisphere mid and high latitude sector (NH,c), respectively. Panel b): Normalized Logarithmic Score versus lead time $\tau$, for neural networks trained with $(T_F, Z_{NH}, S_F)$ predictors (blue) and $(T_F, Z_{NAE}, S_F)$ predictors (red), and with datasets of length 7,200 years (plain lines), 800 years (dashed lines) and 100 years (dashed-dotted lines). The results illustrate the lack of data regime, with very slow convergence of the prediction skill with the dataset length, and with a clear tradeoff between dataset length and size of optimal geographical area for best prediction.}} 
			\end{figure}
			
			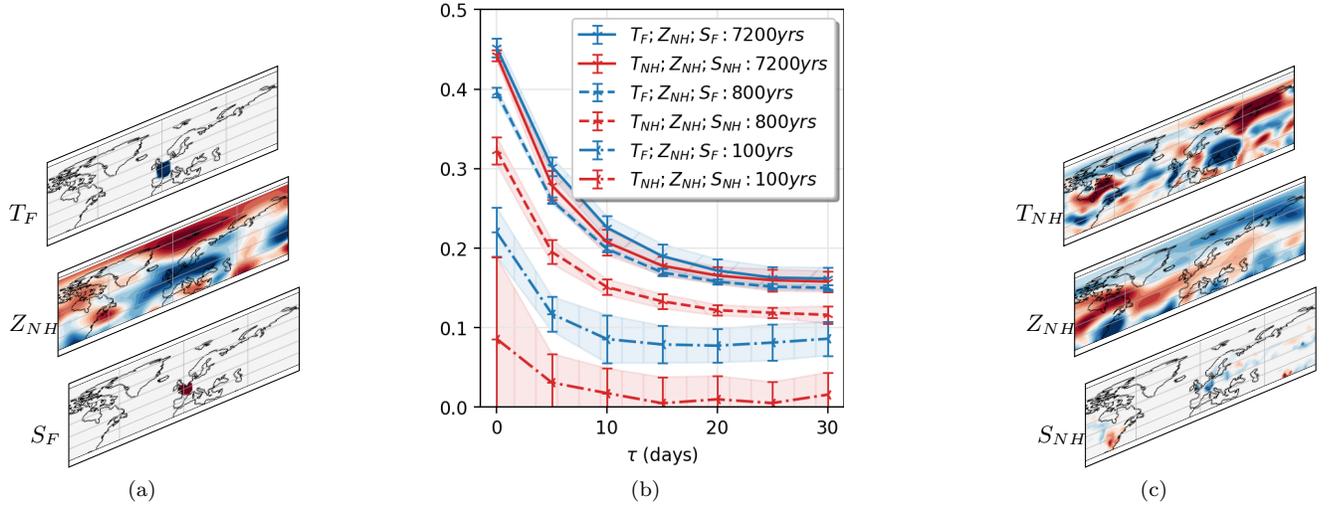
\begin{figure}
				\centering
				\subfloat[\label{fig:fields_inputs}]{%
					\centering
					\tdplotsetmaincoords{60}{50}
					\begin{tikzpicture}[tdplot_main_coords,line join=miter,font=\sffamily,scale=0.8]
						\tikzset{Relu/.style={red, draw=blue, fill=green!20, minimum width=1.5cm,minimum height=0.75cm}}
						\tikzset{ReluFlat/.style={red, draw=blue, fill=yellow!20, minimum width=1.5cm,minimum height=0.75cm}}
						\tikzset{Softmax/.style={red, draw=blue, ultra thick, fill=red!30, minimum width=2cm,minimum height=1cm}}
						\path[tdplot_screen_coords] (-2,0); 
						\pgfmathsetmacro{\xinitialstretch}{4.5} 
						\pgfmathsetmacro{\xstretchchange}{0.25} 
						\edef\Cols{red,red,dgreen,blue,purple,green,yellow} 
						\edef\LstCols{{"red","red","dgreen","blue","purple","green","yellow"}} 
						\pgfmathsetmacro{\yslope}{0.4} 
						\foreach \Col [count=\X starting from 0,remember=\X as \LastX] in \Cols 
						{
							\pgfmathsetmacro{\xstretch}{\xinitialstretch-\xstretchchange*\X} 
							\pgfmathsetmacro{\Xp}{2+2*\X} 
							{\foreach \XX in {0,1,...,\Xp} 
								{
									\pgfmathsetmacro{\pagesep}{0.28} 
									\pgfmathsetmacro{\pagesepchange}{0.05} 
									\pgfmathsetmacro{\pageloc}{\xstretch*\X+1*\pagesep*\XX-1*\pagesepchange*\XX*\X} 
									\ifnum\X=0 
									\pgfmathsetmacro{\pagelocz}{2-2*\XX} 
									\else 
									\pgfmathsetmacro{\pagelocz}{0}
									\fi
									\def\pagelocstart{\xstretch*\X+1*\pagesep*\XX-1*\pagesepchange*\XX*\X-1*\pagesep} 
									\begin{scope}[canvas is yz plane at x=\pageloc]
										\pgfmathtruncatemacro{\fullness}{120-20*\XX} 
										\ifnum\X>0
										\else
										\ifnum\XX=0
										{
											\node (BN-\X) at (-3,2) {$T_F$};
											\draw[fill overzoom image=t2m_input0.png] (-2.5,-.8+\pagelocz) rectangle (2.5,.8+\pagelocz); 
										}
										\fi
										\ifnum\XX=1
										{
											\node (BN-\X) at (-3,0) {$Z_{NH}$};
											\draw[fill overzoom image=zg500_input0.png] (-2.5,-.8+\pagelocz) rectangle (2.5,.8+\pagelocz); 
										}
										\fi
										\ifnum\XX=2
										{
											\node (BN-\X) at (-3,-2) {$S_F$};
											\draw[fill overzoom image=mrso_input0.png] 	(-2.5,-.8+\pagelocz) rectangle (2.5,.8+\pagelocz); 
										}
										\fi
										\fi
									\end{scope}
								}

							}
						}
					\end{tikzpicture}
				}\hfill
				\subfloat[\label{fig:amountofdata}]{%
					\centering
					\includegraphics[width=0.3\textwidth]{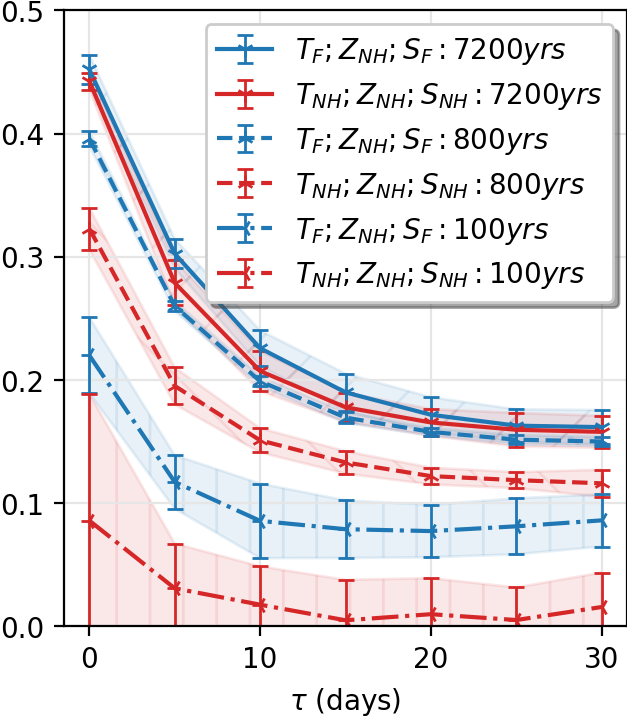}
					
				}\hfill
				\subfloat[\label{fig:full_input}]{%
					\centering
					\tdplotsetmaincoords{60}{50}
					\begin{tikzpicture}[tdplot_main_coords,line join=miter,font=\sffamily,scale=0.8]
						\tikzset{Relu/.style={red, draw=blue, fill=green!20, minimum width=1.5cm,minimum height=0.75cm}}
						\tikzset{ReluFlat/.style={red, draw=blue, fill=yellow!20, minimum width=1.5cm,minimum height=0.75cm}}
						\tikzset{Softmax/.style={red, draw=blue, ultra thick, fill=red!30, minimum width=2cm,minimum height=1cm}}
						\path[tdplot_screen_coords] (-2,0); 
						\pgfmathsetmacro{\xinitialstretch}{4.5} 
						\pgfmathsetmacro{\xstretchchange}{0.25} 
						\edef\Cols{red,red,dgreen,blue,purple,green,yellow} 
						\edef\LstCols{{"red","red","dgreen","blue","purple","green","yellow"}} 
						\pgfmathsetmacro{\yslope}{0.4} 
						\foreach \Col [count=\X starting from 0,remember=\X as \LastX] in \Cols 
						{
							\pgfmathsetmacro{\xstretch}{\xinitialstretch-\xstretchchange*\X} 
							\pgfmathsetmacro{\Xp}{2+2*\X} 
							{\foreach \XX in {0,1,...,\Xp} 
								{
									\pgfmathsetmacro{\pagesep}{0.28} 
									\pgfmathsetmacro{\pagesepchange}{0.05} 
									\pgfmathsetmacro{\pageloc}{\xstretch*\X+1*\pagesep*\XX-1*\pagesepchange*\XX*\X} 
									\ifnum\X=0 
									\pgfmathsetmacro{\pagelocz}{2-2*\XX} 
									\else 
									\pgfmathsetmacro{\pagelocz}{0}
									\fi
									\def\pagelocstart{\xstretch*\X+1*\pagesep*\XX-1*\pagesepchange*\XX*\X-1*\pagesep} 
									\begin{scope}[canvas is yz plane at x=\pageloc]
										\pgfmathtruncatemacro{\fullness}{120-20*\XX} 
										\ifnum\X>0
										\else
										\ifnum\XX=0
										{
											\node (BN-\X) at (-3,2) {$T_{NH}$};
											\draw[fill overzoom image=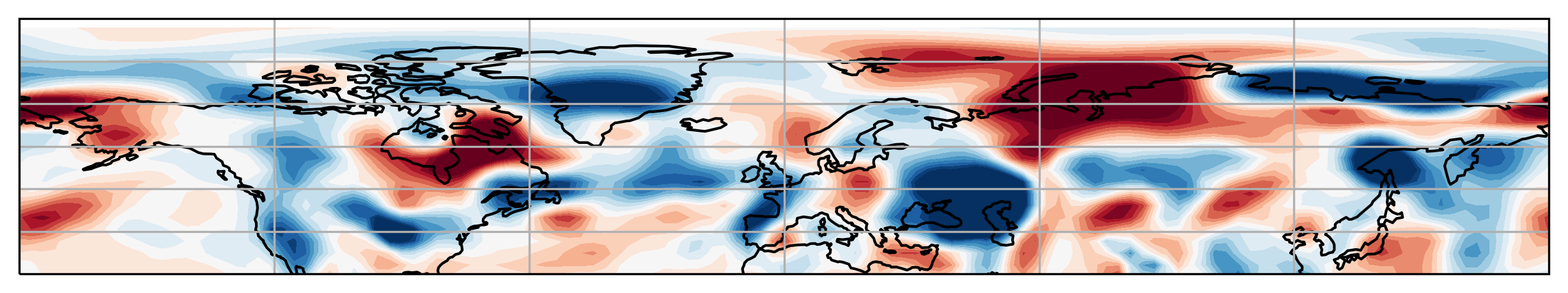] (-2.5,-.8+\pagelocz) rectangle (2.5,.8+\pagelocz); 
										}
										\fi
										\ifnum\XX=1
										{
											\node (BN-\X) at (-3,0) {$Z_{NH}$};
											\draw[fill overzoom image=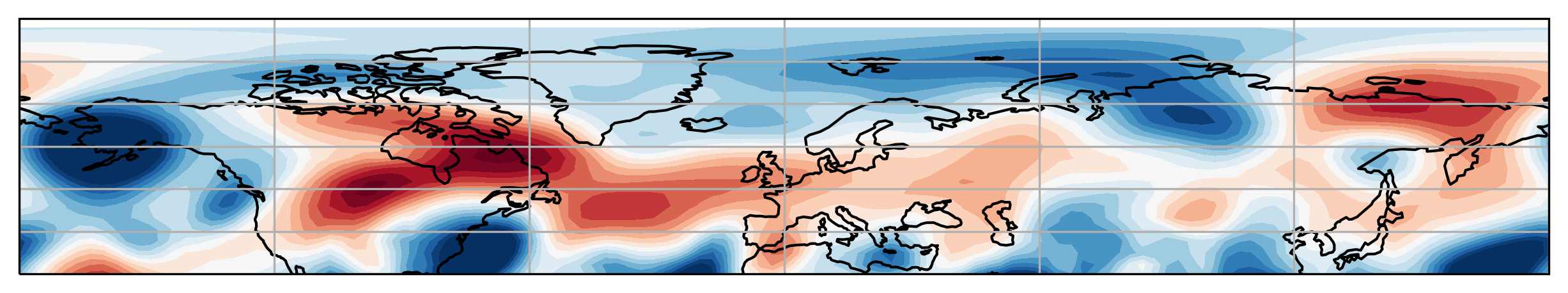] (-2.5,-.8+\pagelocz) rectangle (2.5,.8+\pagelocz); 
										}
										\fi
										\ifnum\XX=2
										{
											\node (BN-\X) at (-3,-2) {$S_{NH}$};
											\draw[fill overzoom image=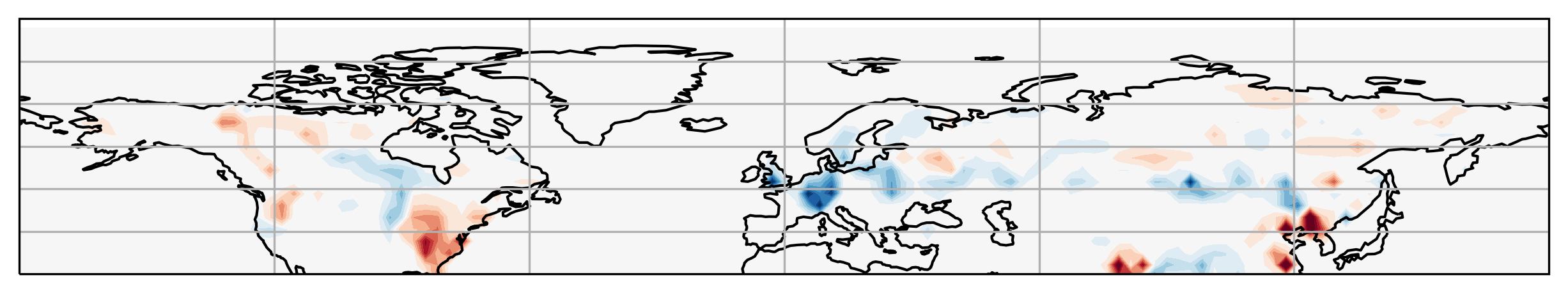] (-2.5,-.8+\pagelocz) rectangle (2.5,.8+\pagelocz); 
										}
										\fi
										\fi
									\end{scope}
								}

							}
						}
					\end{tikzpicture}
				}
				\caption{Prediction skills versus dataset lengths and optimal geographical area for the 2-meter temperature and soil moisture. Panels a) and c) show typical predictor fields, either masked over some restricted area (panel a), or over the whole mid and high latitude Northern Hemisphere (panel c). Panel b): Normalized Logarithmic Score versus lead time $\tau$, for neural networks trained with $(T_F, Z_{NH}, S_F)$ predictors (blue) and $(T_{NH}, Z_{NH}, S_{NH})$ predictors (red), and with datasets of length 7,200 years (plain lines), 800 years (dashed lines) and 100 years (dashed-dotted lines). The results illustrate the lack of data regime, with very slow convergence of the prediction skill with the dataset length, and with a clear tradeoff between dataset length and size of optimal geographical area for best prediction. The optimal area for 2-meter temperature and soil moisture is the local one (France area). }
				\label{fig:data_reduct_masking}
			\end{figure}

			
			The reanalysis datasets~\cite{ecmwf2020} assimilate all available observations with the laws of physics embedded in the weather models, and thus offer the most precise available approximation of the real state of the atmosphere. They are, however, only available during the last 70 years, at most. Is such a short dataset long enough to make reliable prediction using neural networks? One of the key goals of this work is to understand the effect of dataset lengths on the probabilistic predictions that can be issued by neural networks. This is an important question, because many practical applications of  neural networks in atmosphere and climate sciences currently use reanalysis datasets for both training and validation. 
			
			Within our PlaSim model, we now study the effect of reduction of the training set on the prediction skill. We use the prediction skill with a neural network trained on 7,200  years of data presented in section~\ref{field_comparison} as a benchmark. From the PlaSim dataset, we extract two training subsets of shorter year span: 100 years and 800 years. For both cases, we estimate the skill on a validation set that contains the complement of the full 8,000-year long dataset\footnote{In case of 800 years we actually invert validation and training sets that were taken for the benchmark case and perform 10-fold cross validation. In case of 100 years we only sample 10 representative training sets of 100 years and validate on the remainder}. For these experiments, we chose the predictors $(T_F,Z_{NH},S_F)$ which have been proven optimal in section~\ref{field_comparison}, when using a 7,200 year training dataset.
			
			The results are shown by the blue curves on Figure~\ref{fig:datasize-b}. The conclusion is that reducing the dataset up to 100 years has severe consequences for the prediction skill, with a Normalized Logarithmic Score $NLS$ (see equation~\ref{eq:NSS}) nearly halved compared to the benchmark obtained with a 7,200 year training set. We stress that for the 100-year training set, even the plateau skill, corresponding to the effect of soil moisture only, is not correctly predicted. When using a 800-year training dataset, the prediction skill is still quite significantly lower than when using a 7,200-year one. However, the difference of $NLS$ is now of the order of about 10$\%$ at most. This suggests that the convergence of the skill with the dataset length probably occurs on the order of a few thousands to a few tens of thousands of years, if one uses only the three predictor fields $(T_F,Z_{NH},S_F)$. We thus conclude that as long as the source for training of neural network contains only few centuries or even millennia this results in the regime of lack of data, which consequently implies a regime of drastic lack of data when using reanalysis datasets.\\
			
			In such a regime it is customary for machine learning applications that there exists a tradeoff between the dataset length and the complexity of the predictors. Indeed the amount of requested data for optimal training tends to increase when more features are included, in other words greater variety of predictors may lead to overfitting. We now study this tradeoff, as another and complementary manifestation of the regime of lack of data for  neural networks applied to extreme heatwaves. 
			
			To this end we train a neural network with the predictor set $(T_F,Z_{NAE},S_F)$, where the 500 hPa geopotential height information is used only on the North Atlantic and European area: $Z_{NAE}$. We will compare its skill to the benchmark one $(T_F,Z_{NH},S_F)$ that uses the 500 hPa geopotential height on the whole Northern Hemisphere mid and high latitude. Dynamically, the information on the North Atlantic and European sector is more important for France heatwave than the information on the rest of the Northern Hemisphere (see for instance~\cite{yiou2014anawege,yiou2019stochastic}). However, we have recently demonstrated that extreme heatwaves are associated with hemispheric teleconnection patterns~\cite{Ragone18}. It is thus likely that the rest of the Northern Hemisphere should contain useful complementary information which might be more difficult to learn. We then expect that if we have sufficiently long training datasets, the neural network should have a better skill with the complete field $(T_F,Z_{NH},S_F)$. Otherwise the predictor will turn out to be too complex which would result in the degradation of the Normalized Logarithmic Score. 
			
			The red curve on Figure~\ref{fig:datasize-b} presents the result for the predictor set $(T_F,Z_{NAE},S_F)$, to be compared with the benchmark curve with $(T_F,Z_{NH},S_F)$. Comparing the plain red and blue curve, we see that with a 7,200-year training dataset, the $(T_F,Z_{NH},S_F)$ predictor set is indeed the optimal one. With a 7,200 year training dataset the neural  network is actually able to extract the supplementary information beyond the one which is contained in the North Atlantic and European area. The improvement is significant with increase of the Normalized Logarithmic Score up to 10\%, which is important. 
			
			Comparing now the dashed red and blue curves, for the case with 800 long training dataset, one clearly sees the same pattern, although with a smaller improvement when comparing the predictor for the complete field $(T_F,Z_{NH},S_F)$ and the one with the incomplete one $(T_F,Z_{NAE},S_F)$. However, using only 100 years for training, the dashed-dotted line features an opposite conclusion. The training experiment with the incomplete field, on the North-Atlantic Europe sector, gives a better skill than the training with the complete field. The interpretation we give is that a 100 year-long training set is not complete enough to deal with the complexity of the predictor defined on a larger area. This is a manifestation of the tradeoff between dataset length and predictor complexity in a regime of lack of data. This confirms our qualitative prediction and makes it quantitative.
			
			We complement this study of the tradeoff between predictor complexity and dataset length in a regime of lack of data, by discussing the cases of soil moisture and 2-meter temperature. For those fields, the situation is different because it might be clear on  physical grounds that these two fields are relavent mainly locally, close to the heatwave area. Figure~\ref{fig:amountofdata} features the same benchmark blue curve as the one on Figure~\ref{fig:datasize-b}: the prediction skill for a neural network trained with the optimal predictors $(T_F,Z_{NH},S_F)$. It also shows the prediction skill for a neural network trained with the predictors $(T_{NH},Z_{NH},S_{NH})$, where now both the soil moisture and temperature fields are used on the full Northern Hemisphere mid and high latitude sector. The results clearly show that this prediction with hemispheric temperature and soil moisture is systematically worse than the one with local predictors. This confirms that the optimal area is the local France one, for these two fields. Thus complexifying fields with parts that contain no relevant information and provide essentially noise might be manageable with neural networks trained on huge datasets, but it is a problem in a regime of lack of data. Figure~\ref{fig:datasize-b} also shows that the degradation of the score, in relative terms, is lower when the amount of data is increased, in agreement with our interpretation. 
			
			Our third conclusion is that learning the probabilities of extreme heatwaves with neural networks clearly takes place in a regime of drastic lack of data. Several hundreds or even thousands of years would be needed for optimal prediction, even when using only two or three representative fields as predictors. Trying to use more fields, for instance more information about the vertical structure of the geopotential height, or for example information related to the temporal development of the dynamical fields, most probably requires even longer training datasets. Our results clearly show that with a 7,200-year training dataset, the skill is at best only very marginally improved when using more fields. Moreover, in this regime of lack of data, there is a tradeoff between dataset length and predictor complexity. For instance, for predicting extreme heatwaves over France, benefiting from hemispheric information beyond the North Atlantic and Europe sector requires at least several hundreds of years of learning datasets.
			
			
			


			\subsection{Physical insight, interpretability of neural network predictions, and committor function composite maps}
			\begin{figure*}
				\centering
				\subfloat[\label{fig:q1}]{%
					\centering
					\includegraphics[width=8cm]{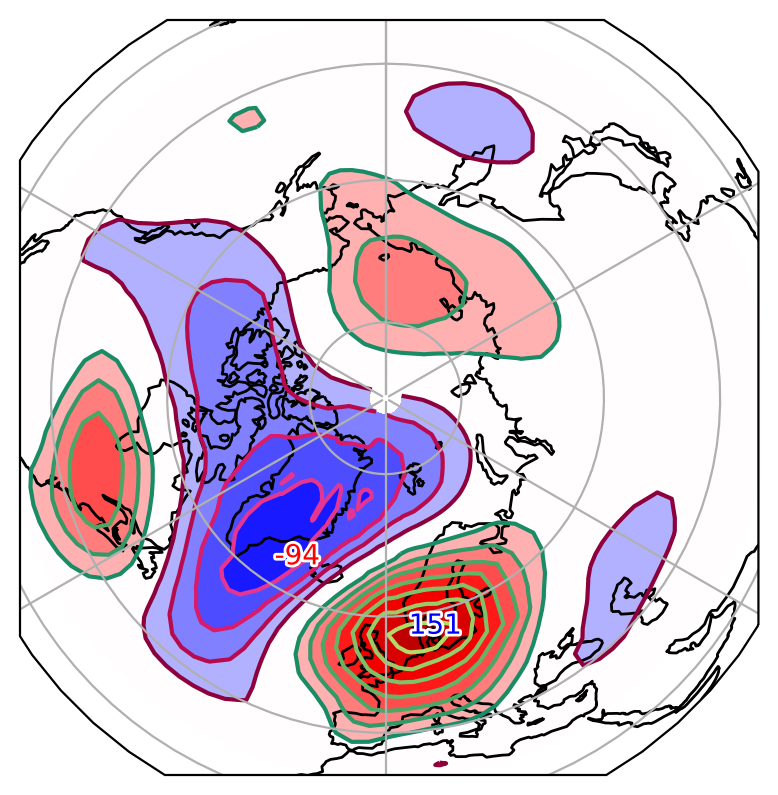}
				}
				\subfloat[\label{fig:q2}]{%
					\centering
					\includegraphics[width=8cm]{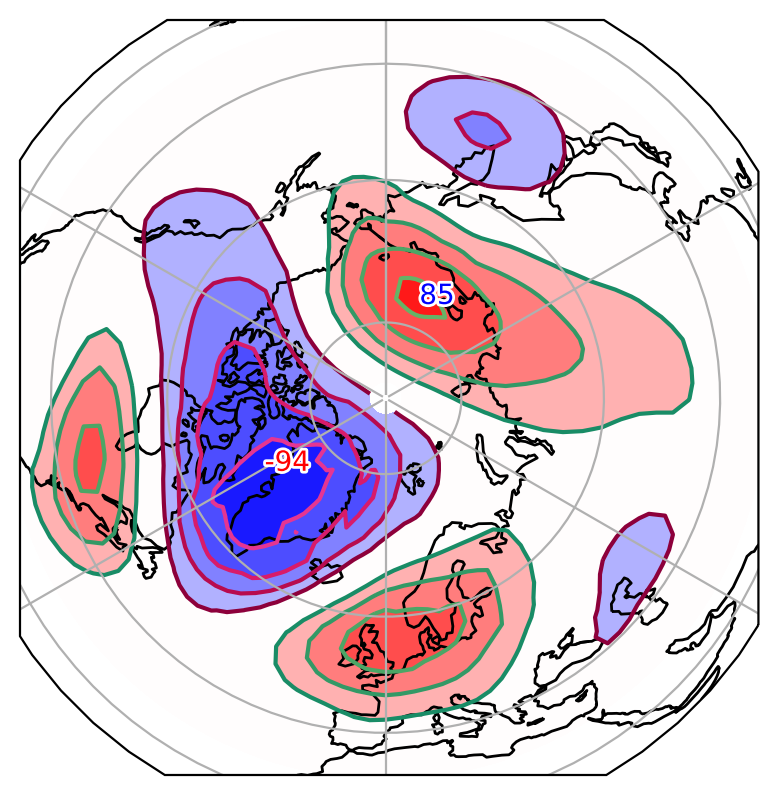}
				}\hfill
				\subfloat[\label{fig:q3}]{%
					\centering
					\includegraphics[width=8cm]{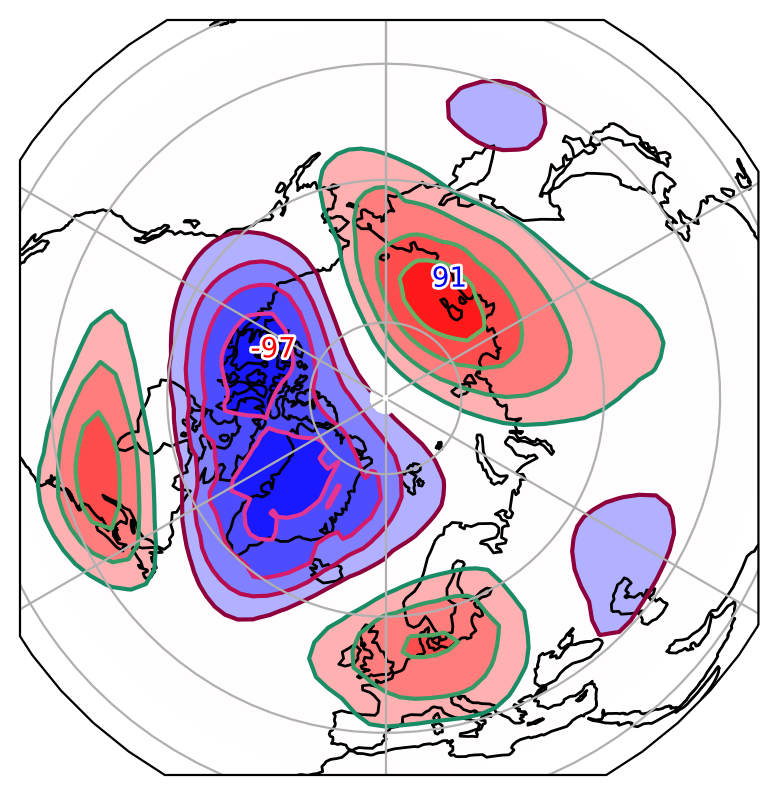}
				}
				\subfloat[\label{fig:q4}]{%
					\centering
					\includegraphics[width=8cm]{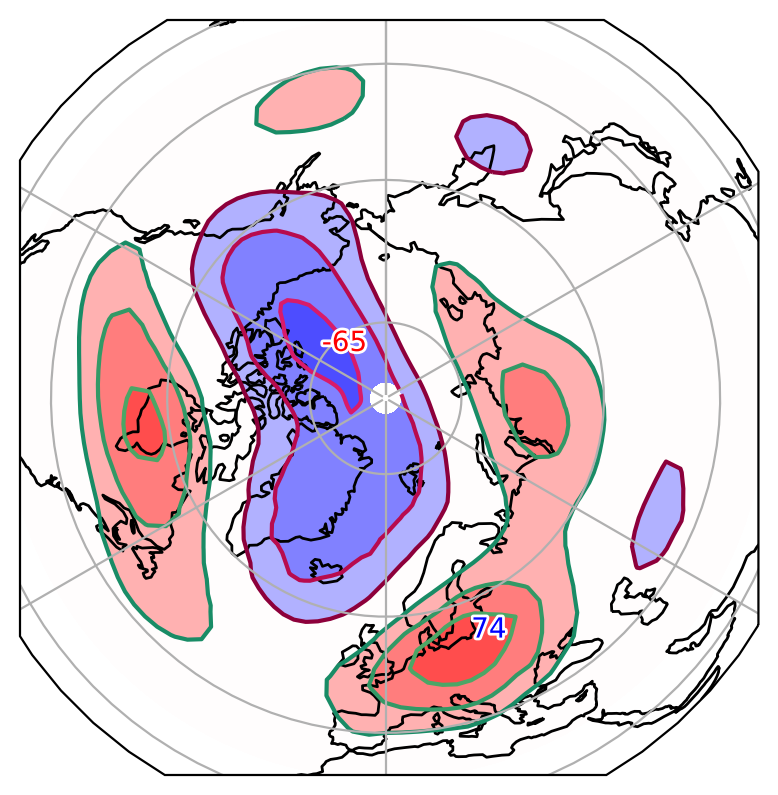}
				}
				\caption{Composites of the 500 hPa geopotential height maps $Z_{NH}$ (in meters), conditioned on committor values $p$ above the $99.9$ percentile, at different lead time $\tau$. Panel a), b), c) and d) with values of $\tau=0,5,10$ and 15, respectively. The regions of positive anomalies are indicated in red, while the negative anomalies are indicated in blue (we are using seismic colormap). The iso-lines are separated by a value of $20$ meters. The maximum geopotential height anomaly is indicated with a number colored in blue, while the minimum value is colored in red.}
				\label{fig:committor_composites}
			\end{figure*}
			
			What has the neural network actually learned?  The interpretability of neural network predictions is a pervasive question when they are applied in physical sciences. In this section we propose a basic approach for visualizing committor function. With this aim, we plot composite maps of the 500 hPa geopotential height, conditioned on very large values of the committor function. 
			
			The  neural network output is the probability $p(\x)$ to observe a heatwave $\tau$ days from now, given that we observe today the predictor field $\x$. $p(\x)$ is a function over the set of all the possible predictor states. This function, called a committor function, is therefore extremely complex and impossible to visualize for high dimensional spaces, by contrast with committor functions for simple dynamical systems~\cite{lucente2020machine}. In order to get insights about some very specific behaviors of this function, we will try to look at a single property: how do the fields that give very large values of $p(\x)$ look like? Equivalently, we will consider the fields that the neural network ranks as the most likely to produce a heatwave, and look at the average of their corresponding 500 hPa geopotential height field. 
			
			In order to simplify the discussion, we consider a neural network trained on the hemispheric 500 hPa geopotential height $Z_{NH}$ only. Once it has been optimized using the training set, the neural network can associate to any other field, for instance, in the validation set, the estimated probability $p$ that this field will lead to a heatwave. We select all the events in each of the validation sets which are above the $99.9$ percentile in the distribution of the committor values. For example, for $\tau=0$ those are all the events with $p>0.68$. Notice that according to our protocol we have 10 folds of train-validation split (section~\ref{sec:training}). This means that we can associate a committor value with each day of the full 8000 year long set and no occasion are we evaluating the committor in the training set.  We then compute the average of the 500 hPa geopotential height maps, conditioned on having $p$ values above the $99.9$ percentile, for the entire 8000 year long set. The resulting composite maps displayed on Figure~\ref{fig:committor_composites} reflect averaged properties of the fields which are most likely to lead to a heatwave, according to the neural network prediction. This operation can be repeated for different values of lead times $\tau$, which may have smaller values of $p$ threshold, simply because the neural network becomes less certain for larger values of $\tau$
			
			 For $\tau=0$, on Figure~\ref{fig:q1}, we observe a tripole structure for the geopotential height anomalies. First of all, we see an anticyclonic anomaly over Europe, which is expected, given that heatwaves in summer are associated with anticyclonic anomalies. We also see a cyclonic anomaly over Greenland and the Arctic area and two other anticyclones over Eastern North America and Northern Siberia. For this composite, the anticyclonic anomaly over Europe is extremely strong: the maximum value of the composite average of the 500 hPa geopotential height anomaly has a maximum value of 151 meters (see Figure~\ref{fig:q1}), to be compared with a typical maximal 500 hPa geopotential height anomaly over Europe of order of 120 meters (see for instance the snapshot on Figure~\ref{fig:synoptic}) and a variance for the climatology of the 500 hPa geopotential height anomalies at midlatitude of order of 60 meters. Obtaining such a large value for a composite average, means that all the fields in the composite have a systematic stronger than usual anomaly over Europe with a coherent pattern. Similarly the cyclonic anomaly over Greenland is very strong, with a minimum value of the averaged 500 hPa geopotential height anomaly of -94 meters, to be compared with typical minimal 500 hPa geopotential height anomalies over the Greenland-Artic area of order of -200 meters (see for instance the snapshot on Figure~\ref{fig:synoptic}) and a variance for typical 500 hPa geopotential height anomalies of order of 90 meters at high latitudes. This also points to a very coherent and systematic pattern over Greenland. The two other anticyclonic anomalies over Eastern North America and Northern Siberia have weaker values, of order of 40 to 60 meters, which are still comparable to the variance of the 500 hPa geopotential height, thus showing a relatively strong coherence. The coherence of the overall pattern can also be assessed by comparing those values to the standard deviations within the composite set itself, which are of order of 40 meters in midlatitudes and 60 to 90 meters in the Arctic area. All those comparisons point to fairly coherent and robust patterns superimposed with fluctuations of the order of the standard deviation. 
			
			The overall pattern is a clear mode 3 pattern, with an overall shift of the cyclonic anomalies poleward and of the anticyclonic anomalies equatorward. This structure is much reminiscent of the wavenumber 3 extreme teleconnection observed for European heatwaves~\cite{Ragone18} and has been interpreted as related to Rossby waves with wave-number 3, and phase speed close to zero, leading to a long-lasting quasistationary pattern. This result suggests that the recognition of this wavenumber quasistationary pattern might be key for the neural network prediction skill. 
			
			Let us now look at different values of the time lag $\tau$. The three other panels of Figure~\ref{fig:committor_composites} show the composite 500 hPa geopotential height maps, conditioned on $p$ values above the $99.9$ percentile, for $\tau=5,10$ and 15, respectively.  The four patterns look surprisingly similar, whatever the value of $\tau$, with a consistent wave number 3 pattern, poleward shift of the anticyclonic anomalies and equatorward shift of the cyclonic ones. This result suggests that the long-term prediction skill of the neural network might also be associated with this quasistationnary pattern. Following this remark a natural hypothesis would be that the long-term skill of the  neural network might be related to the probability of this pattern to stay quasistationnary for a long enough period. Testing this very interesting hypothesis is however beyond the capabilities of the approach described in this paper, and will be considered in future works. 
			
			In addition to the strong analogies, we note that the 4 patterns are slightly modified when changing $\tau$. For $\tau=5$ the anticyclonic anomaly over Northern Asia is stronger and larger, and the anticyclone over Europe is less intense. This tendency is even more pronounced for $\tau=10$. For $\tau=15$, the wave number 3 pattern turns to a tripolar structure.  
			
			Plotting composite 500 hPa geopotential height maps, conditioned on very large values of the committor function, gives most probably only a limited view of what the neural network might have learned. Trying to interpret the neural network results with an averaged quantity (composite) only, is very limited from the point of view of a stochastic interpretation of the prediction. However these composite averages already clearly show very interesting teleconnections associated with previously discussed wave-number 3 patterns. This opens questions for more detailed future analysis of the dynamical features which are important for prediction, their probabilities, and the capabilities of the  neural network to identify them. Those future analysis will also consider tools for physical interpretability of machine learning forecasting.
			
			\subsection{How to ensure continuity of the committor function when the lead time is changed 
				and how to accelerate the training stage}\label{sec:transfer}
			
			\begin{figure*}
				\subfloat[\label{fig:timeevolution}]{%
					\centering
					\includegraphics[width=8.6 cm]{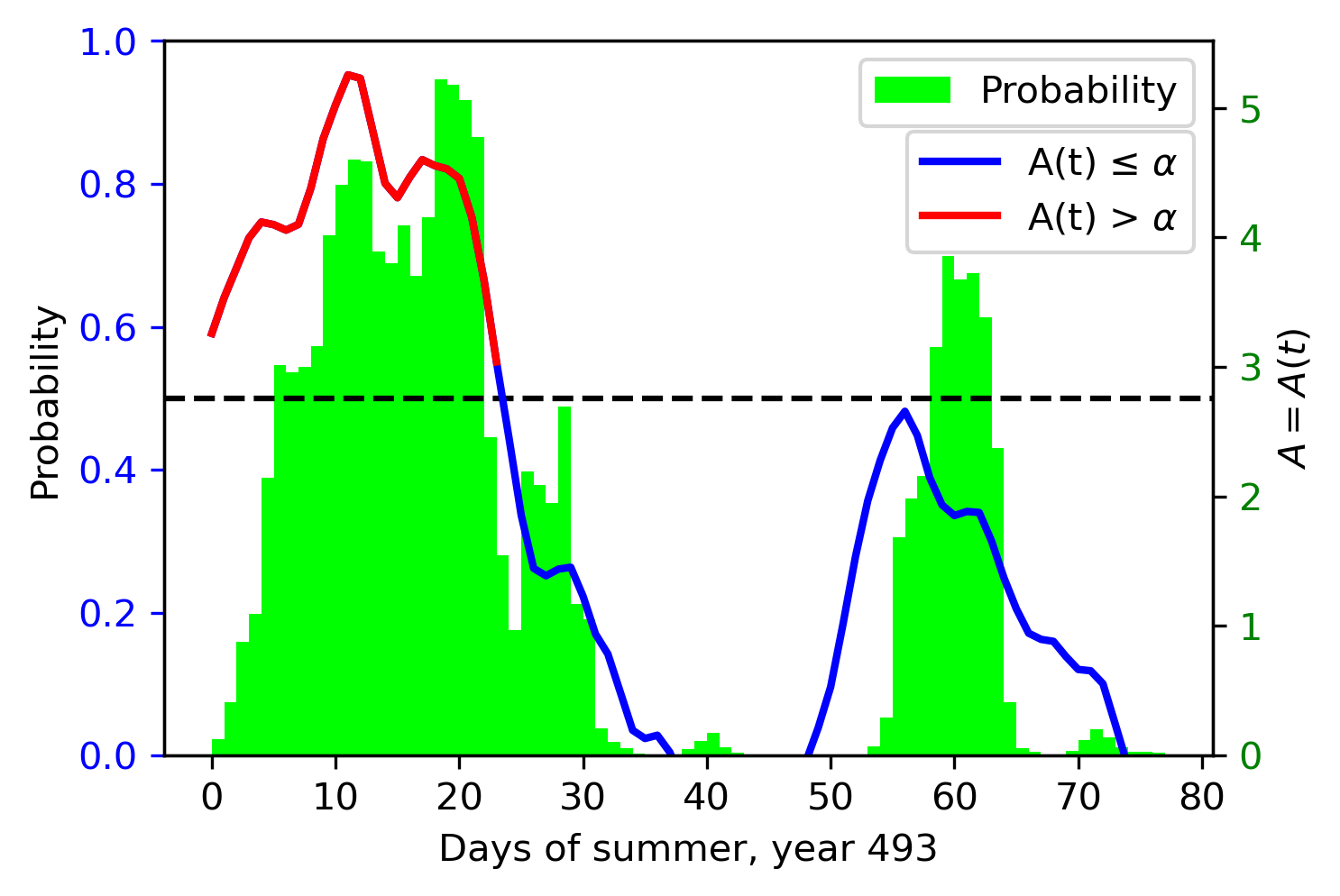}
				}\hfill
				\centering
				\subfloat[\label{fig:smoothness}]{%
					\centering
					\includegraphics[width=8.6 cm]{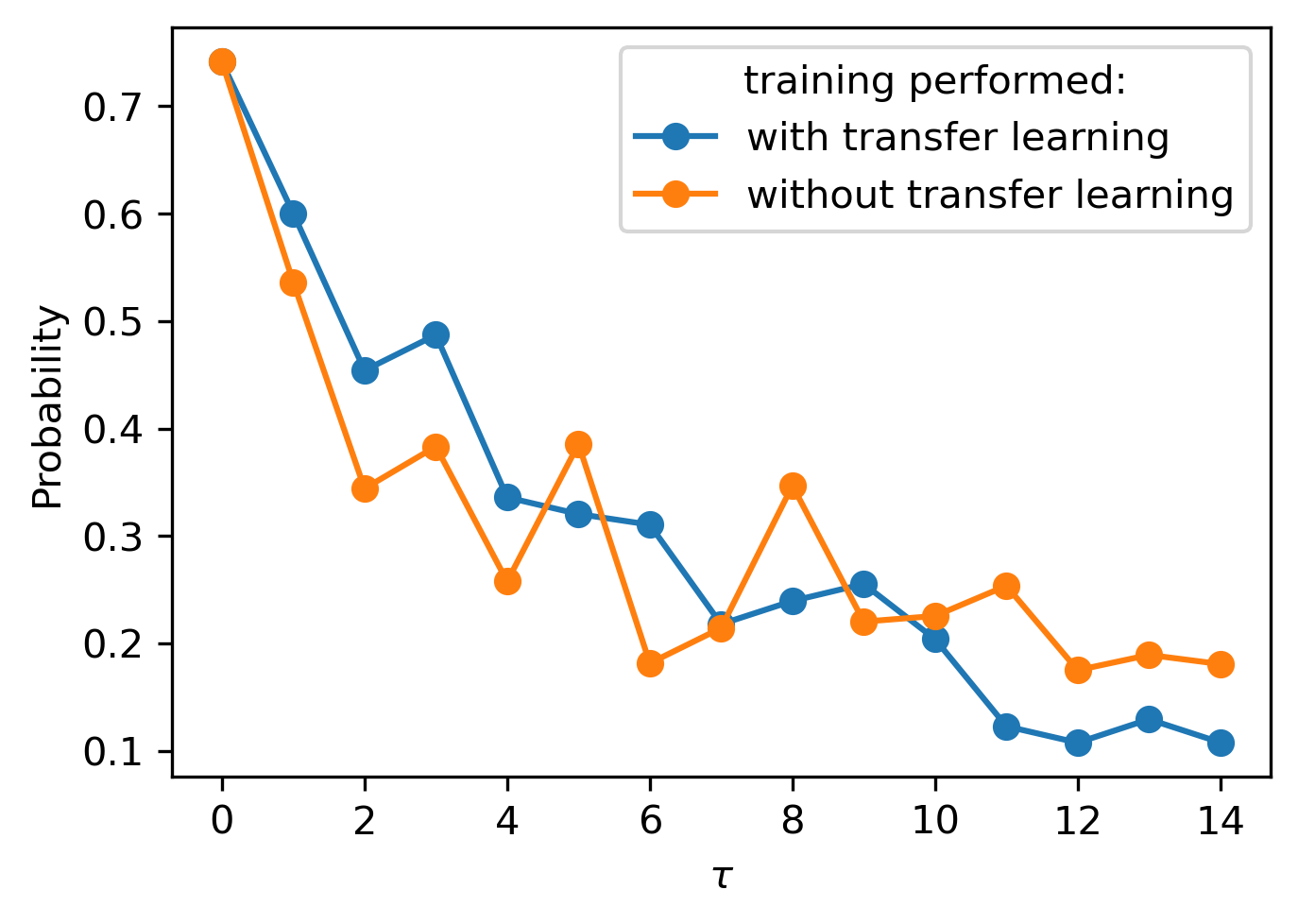}
				}
				\subfloat[\label{fig:stddiffqtau}]{%
					\centering
					\includegraphics[width=8.6 cm]{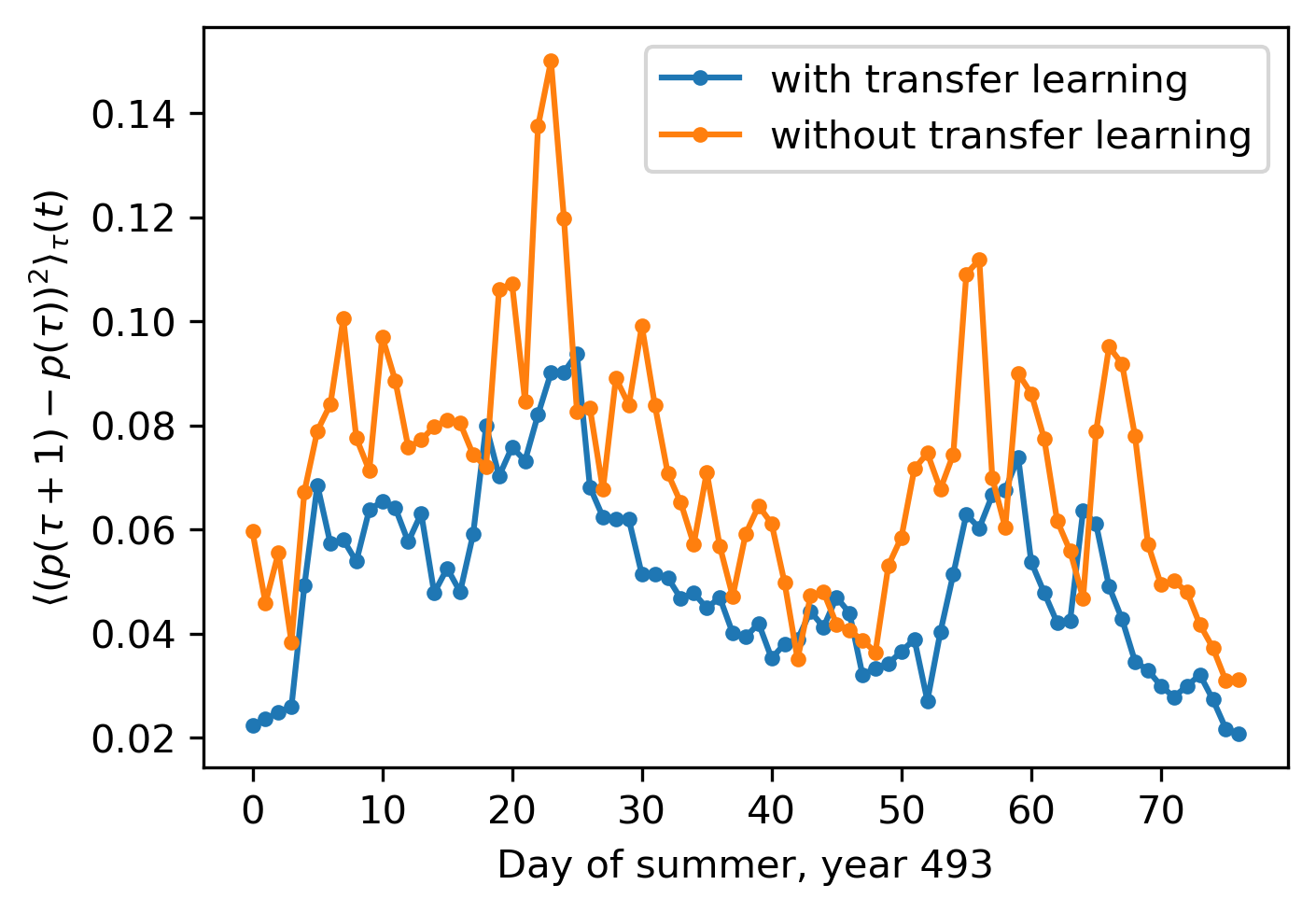}
				}
				\caption{{(a) {{Illustration of the $A(t)$(blue/red line) and $p\left[X(t),0\right]$ (in green) trajectories during summer of a specific year; red line corresponds to moments where the actual $A = A(t)>\alpha$ (see equation~\eqref{timeaveraged}), above 95 percentile threshold, while blue line denotes that $A = A(t)\le\alpha$. The filled green segments show the probability predicted by the neural network trained on $T_F, Z_{NH}, S_F$. This plot illustrates the general correlation between $A = A(t)\le\alpha$ and $T_F, Z_{NH}, S_F$. 
						(b) Evolution of the committor function along a trajectory $p\left[X(t^\star-\tau),\tau\right]$ as a function of lead time $\tau$, for a prediction of a potential heatwave at a prescribed physical time $t^\star$. The orange points show the committor learned independently for each $\tau$, while the blue ones are obtained with transfer learning from one $\tau$ to the next.
								(c)  Display of $\sigma_\tau(t)$, as defined in equation~\eqref{gradvariancefocommittor}, for the same year as in (b), for the method with transfer learning (in blue) and without (in orange). We see that the latter is almost always above. This illustrates that variance between subsequent points is larger when transfer learning is not applied.}}}}
			\end{figure*}

			The proposed neural network predicts the committor function: the probability $p(\x,\tau)$ to observe a heatwave $\tau$ days from now, given that we observe today the predictor field $\x$. 
			First, we display in Figure~\ref{fig:timeevolution} a trace of this committor function (in green) at $\tau=0$, for the summer of a randomly chosen year, compared to the actual realization of temperature anomaly $A$. One sees that committor function and actual events are correlated, yet not identical, as expected {since the committor is only a probability of having a heatwave and the neural network is only trained on discrete labels and is not provided the full information $A(t)$}.  
			
			Up until now, we have studied  some aspects of how 
			{the total prediction skill evaluated on validation sets depended on $\tau$ }. In this sub-section, we {are rather interested in how committor function varies along the trajectory (passage of $t$ physical time in the simulation) while fixing a specific event at time $t^\star$. Thus we must vary simultaneously $\tau = t^\star - t$. Of particular interest is a smoothness property of the committor as we vary $t$ while fixing $t^\star$.  }

			This smoothness property could be understood as a consistency of the prediction through time. A lack of smoothness, in a risk prevention context, would mean that the prediction would highly fluctuate from one day to another. Besides the fact that this would probably be the sign of some deficiency in the prediction, this might also be detrimental from a communication point of view, and create concerns and disbelief among the user of the information. On more scientific grounds, if the prediction $p$ is used as an input for another computation or algorithm, the consistency and smoothness properties might also be very important, both theoretically and practically. 
			
			We demonstrate in this section that transfer learning can be used to address this issue. Transfer learning is used extensively in deep learning, where it allows dramatic reduction of the training time, and improvement of the skill, by using networks pre-trained on more general large datasets~\cite{Jiallin10,Waseem17}. It has also been used for several climate and weather applications, for instance for ENSO prediction~\cite{Ham} where the authors have pre-trained the network on CMIP model outputs, prior to applying it to reanalysis datasets. When looking for smoothness properties with respect to $\tau$, an alternative could be to train on several lag times at the same time, see for instance~\cite{weyn21} and references therein.\\
			
			To study the $\tau$ dependence of $p$, one could either fix the state $\x$, or rather follow the evolution of state $\x$ with time. In this section, we make this second choice. We thus fix a time $t^\star$ corresponding to the potential start of a heatwave, and we study how $p\left[X(t^\star-\tau),\tau\right]$ depends on $\tau$. When $\tau$ decreases, the prediction is made closer in time to the start of a potential heatwave. We then expect the event to be more predictable, as $\tau$ decreases. There is no reason to expect a monotonic evolution. However, in general, we expect $p\left[X(t^\star-\tau),\tau\right]$ to be a smooth function of $\tau$. {Since we are working with discrete daily data the highest resolution we can achieve is obviously $\Delta\tau = 1$ day. Thus smoothness must be understood in colloquial terms rather than a strict mathematical definition. Below we give a specific example.}
			
			We first train the network in an independent way for different values of $\tau$, with the reference predictors $(T_F,Z_{NH},S_F)$. For illustration purposes, we choose a specific year and a value of $t^\star$ such that it corresponds to the strong heatwave at $t^\star$. The orange curve on Figure~\ref{fig:smoothness} shows the evolution of the prediction $p\left[X(t^\star-\tau),\tau\right]$ with $\tau$. One sees that the prediction is relatively consistent over time, when $\tau$ decreases. However we observe some fluctuations, from one day to the next, of the order of 10\% to 30\% of the predicted probability. The level of these fluctuations is higher for intermediate values of $\tau$, between 2 to 10 days.
			
			In order to reduce these fluctuations and to improve time consistency and smoothness of the prediction, we adopt a transfer learning strategy. The main idea is to initialize the weights of the neural network for the model at a given lead time $\tau$, based on the trained model at a previous lead time $\tau-1$. The heuristic idea is that the corresponding change in $X(\tau)$ is not so large and already contains very good information for the prediction at the next time step. Note that this also allows to drastically reduce the training time: early stopping of the training is typically necessary only after 5 epochs, as opposed to 40 or more when starting from random initialization. The reference is the blue curve on Figure~\ref{fig:stddiffqtau}.
			In addition, we tested the effect of this transfer learning strategy on the overall prediction skill but we have seen no significant improvement or degradation. This suggests a hypothesis that we have reached the capacity of the network to learn the extreme events. 
			
			{For quantifying the reduction in the fluctuations of $p$, we introduce a smoothness metric. From the discrete series of $p$, we compute the forward difference of the committor at successive $\tau$:
				\begin{equation}
					\Delta_\tau (t) := p\left[X(t-\tau-1),\tau+1\right] - p\left[X(t-\tau),\tau\right],
				\end{equation}
				which is a function $t$. The smoothness metric consists in computing standard deviation of $\Delta_\tau(t)$ for all $\tau\in\{0,...,14\}$ days:
				\begin{equation}\label{gradvariancefocommittor}
					\sigma^2_\tau(t):=\langle \Delta_\tau^2 (t) \rangle_\tau - \langle \Delta_\tau (t) \rangle_\tau^2
				\end{equation}
				where the brackets $\langle \cdots \rangle_\tau$ denote the average over the subscript parameter $\tau$. 
				Figure~\ref{fig:stddiffqtau}  compares $\sigma_\tau(t)$ in cases with and without transfer learning; it clearly demonstrates that the former tends to have smoother committor function w.r.T.~$\tau$. 
				The level of fluctuations from one lead time to the next one has been reduced by a large factor, when compared to the orange curve without transfer learning.
				We can also apply this quantitative measure of smoothness to the sequence of all days $t$ in each validation set of which we have 10 folds, as described in section~\ref{sec:training}. Since the sequence now also consists of all times $t$, we average on all $t$ and $\tau$, denoted as $\sigma_{\tau,t}$ which is a scalar. This results in the following values with transfer learning $\sigma_{\tau,t }= 2.59 \pm 0.07 \times 10^{-2}$ and without $\sigma_{\tau,t }^\prime = 3.82 \pm 0.09 \times 10^{-2}$. 
				The difference is almost 50\%. This quantitative measure allows us to conclude that transfer learning improves dramatically the time consistency and the smoothness of the extreme heatwave prediction, while, independently, reducing the training computational time.}

			\subsection{Robustness of the learning protocol with respect to the undersampling strategy, the level of rarity, and the neural network architecture}
			
			\subsubsection{How does majority class undersampling affect the prediction skill?}
			\label{app:undersampling}
			
			\begin{figure*}
				\centering
				\includegraphics[width=8.6 cm]{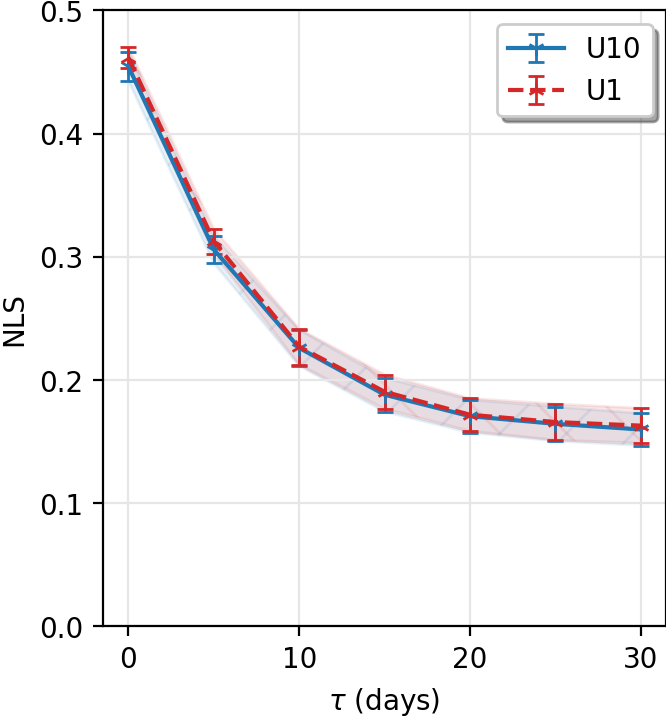}
				\caption{Normalized Logarithmic Score versus lead time $\tau$ for a neural network trained with the $(T_F, Z_{NH}, S_F)$ predictors, with undersampling with $r=10$ (U10) (blue), and without undersampling (U1) (red). The two skill curves are within error bars of the experiment.}
				\label{fig:Undersampling_Skill}
			\end{figure*}
			
			In this section we discuss the effect of majority class undersampling strategies on the neural network prediction skills. We train the neural network with the reference set of predictors $(T_F, Z_{NH}, S_F)$, which was proven optimal in section~\ref{field_comparison}. We either train the network without undersampling, or with the majority class undersampling taking into account the change of probability measure, as discussed in section~\ref{sec:unbalancedunbiasing} and using equation~(\ref{eq:pT_p}). For this second case, the undersampling rate is $r=10$. Figure~\ref{fig:Undersampling_Skill} shows that the prediction skills are the same, within the error bars. This leads to a conclusion that the prediction skill of the extreme heatwaves considered here is not influenced by the undersampling strategy. 
			
			 Majority class undersampling is however useful, in order to reduce the memory request and the training time by about a factor 10. One may wonder if the similar conclusion can be reached for other types of extremes, which could be the subject of future work.
			
			\subsubsection{Prediction skills for more extreme events}
			\begin{figure*}
				\centering
				\includegraphics[width=8.6 cm]{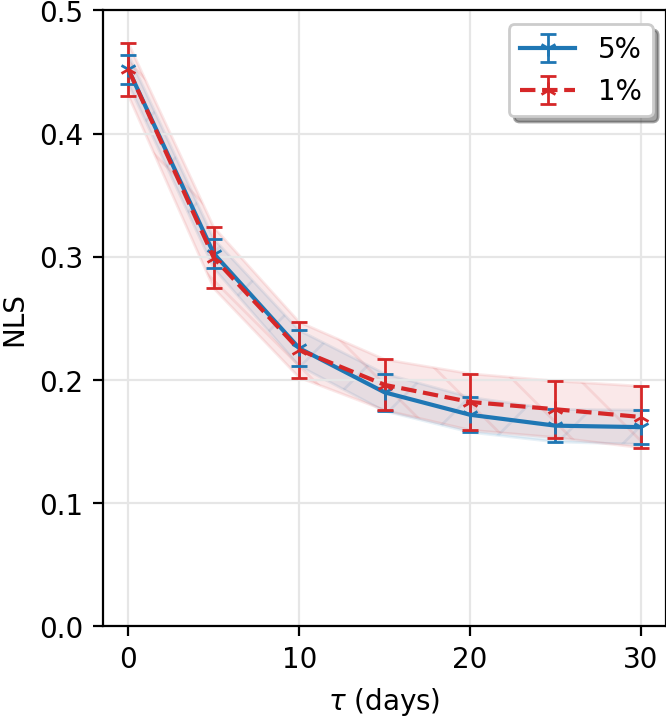}
				\caption{Benchmarks for the optimal Normalized Logarithmic Score obtained by the proposed neural network for 5 percent heatwaves (blue)  and 1 percent heatwaves (red)}
				\label{fig:rarity}
			\end{figure*}
			
			On Figure~\ref{fig:rarity} we present the comparison between predictions of 95 percentile heatwaves (consistent with all the previous analysis) and 99 percentile heatwaves. This corresponds to definitions of heatwaves for large deviations $A(t) > 2.75 {\rm K}$ and much more extreme ones $A(t) > 3.91 {\rm K}$ in the latter case. In other words, the objective field used to make the prediction $X$ is exactly the same ($T_F, Z_{NH}, S_F$) as well as the architecture presented in Fig.~\ref{fig:architecture} but the labels are defined based on two different criteria discussed above. Undersampling rate was chosen as 10 for 95 percentile heatwaves as usual, and 20 for the 99 percentile case. The resulting scores are plotted on Figure~\ref{fig:rarity}. 
			
			At a qualitative level, the normalized logarithmic scores behave similarly for the two cases, with the same decrease of the skill over synoptic time scales, up to a plateau corresponding to the effect of soil moisture. At a more quantitative level, we stress that the nearly equal values of the scores for the two cases is an accident. There is no logical reason to compare directly the quantitive values of the skills for the two experiments. The first reason why they are not comparable is that the normalized logarithmic scores are normalized differently in each case, because of the different base climatological probability. The second reason, is that even with the same climatological probability, there would be no reason to expect two events of different classes to have the same exact real committor value, which is the intrinsic probability we would learn if the learning would be perfect.\\
			
			It is interesting to note that in our previous study~\cite{jacques-dumas22}, majority class undersampling or transfer learning among classes was improving the categorial (0 or 1) prediction of extreme heatwaves, when we were assuming a deterministic relation between predictors and heatwaves. In this new paper, neither majority class undersampling nor transfer learning affect the probabilistic prediction skill, neither positively nor negatively, when we now actually consider the probabilistic nature of the relation between predictors and the heat waves. One might be surprised by these different behaviors. One possibility would be that our learning was not perfect, in one or the other case, either because of lack of data or suboptimal learning protocol. However, we stress that even with a prefect learning, there is no disagreement nor contradiction between these seemingly different results, as what is actually tested for the prediction is of a different nature. 
			
			This rises a very interesting general question. In several previous studies, including ours~\cite{jacques-dumas22}, a categorial test, for instance the Matthews Correlation Coefficient, was used to test a relation between predictors and events which is actually intrinsically probabilistic, and not deterministic. This was logically problematic and should be avoided. Beyond the logical problem, might it be possible that testing a probabilistic relation with a categorial test lead to some practical inconsistencies and divergent conclusions? As we are interested by probabilistic forecast we do not consider further this question.
			
			\subsubsection{Robustness of the results with respect to the neural network architecture}
			
			Throughout the article, the architecture displayed in Figure~\ref{fig:architecture} was consistently used when we refer to the CNN (or neural network) methodology. We have tried other architectures, changing the amount of filters, using additional layers, and other changes for a better optimized network. None were convincing in the present framework. Specifically, we found that deeper CNNs had slightly lower skill, and this is the reason why they are not consider here. 
			
			While we exclusively show results pertaining to stacking the fields, we have also considered combining the fields into separate CNNs which are then concatenated on a single dense layer. The latter approach does not work so well, as already reported in \cite{jacques-dumas22}, and is more difficult to implement. This suggests that stacked architecture is potentially benefiting from local cross-correlations between temperature, soil moisture and geopotential 



\section{Conclusions and perspectives}
\label{sec:conclusion}

\subsection{Probabilistic forecast with machine learning, and other methodological contributions}

In this paper we have advocated a probabilistic approach for the forecast of weather and climate related problems, in particular extreme events, because for chaotic dynamical systems the relation between predictors and the predicted phenomena is intrinsically probabilistic. 
For forecast validation, logarithmic or ignorance score, occasionally used in weather forecast and climate, is directly linked to the cross-entropy skill. The latter is used in many machine learning problems as opposed to, say, Brier score. Through an affine transformation of the logarithmic score we defined the Normalized Logarithmic Score, which has convenient properties to be equal to zero for a forecast based on the climatological frequency, to be positively oriented and to be always lower than one.

We have demonstrated the efficiency of this approach for forecasting long-lasting extreme heatwaves, within a dataset consisting of PlaSim climate model outputs. Using geopotential height, temperature, and soil moisture fields as predictors, we have trained a convolutional neural network to forecasts extreme heatwaves on a validation set. Methodologically, this probabilistic approach extends previous work using machine learning for categorical deterministic prediction of daily~\cite{Chattopadhyay19} or long-lasting heatwaves~\cite{jacques-dumas22}.

At a methodological level, we have also demonstrated the interest of transfer learning in order to improve the temporal consistency and smoothness with time of the prediction. This is a key issue for practical applications related to risk forecast. We have also demonstrated the interest of majority class undersampling and of transfer learning in order to lower the RAM, CPU, and computational time usage during the learning stage of the network.

\subsection{Key general scientific conclusions}

\subsubsection{The lack of data regime for machine learning in weather and climate studies}
\label{sec:con-lack-of-data}

The main scientific message of this work is that training neural networks for predicting large scale features of weather or climate phenomena will most of the time operate in a regime of lack of data. We have demonstrated this clearly in the case of extreme heatwaves. Using subsets of 8,000-year climate output with a data reduction protocol leads to a significant drop in the prediction skill. using three important fields, one at hemispheric scale (500 hPa geopotential height) and two at a local scale (soil moisture and 2-meter temperature). This points to the need of thousands or tens of thousands of years of data for proper convergence, perhaps more if one would like to benefit from the information available in more complementary fields.

The climate model output has some known biases with respect to real fields, but its structure and complexity is most probably the same as the one for reanalysis datasets or real fields. It is likely that obtaining a converged statistical model based on real or reanalysis dataset would require a length of the same order of magnitude, although this cannot be tested directly. This is a drastic constraint given the definitive limitation of historical data. A similar lack of data problem exists for many applications of machine learning for physical and natural sciences, but the lack of observed or reanalysis data is rather severe for studying large scale weather and climate phenomena. In order to circumvent this problem, one will have to find ways to combine model data and reanalysis datasets as discussed in section~\ref{sec:per-lack-of-data}.

This problem is exacerbated when studying extreme events because of their rarity. This is indeed a very important remark. The heatwaves we have studied in this paper, which were defined as the 5\% percentile of summer data with a correlation time of a few days, are events with typical return time of a few years in the studied climate.  Many climate and weather phenomena have return times of a few weeks to a few years, they will equally fall in this lack of data regime. From a point of view of extreme event impact, a return time of a few years is actually not so rare and risk management specialists are interested in much rarer events. \\ 

We have clearly demonstrated that there exists a tradeoff between the length of the dataset and the complexity of the used predictors. For instance, for forecasting extreme heatwaves over France, we showed that the 500 hPa fields contain useful information for improving the prediction skill at the hemispheric scale. However, in order to properly learn part of this information, the neural network needs at least a few hundred years of data. To obtain a larger improvement using hemispheric fields, compared to fields at the scale of North-Atlantic and Europe, actually require thousands of years of data. This tradeoff predictor complexity/dataset length is very natural for machine learning in a context of lack of data, and should be present in most applications of neural networks when studying large scale features of climate or weather data. 

\subsubsection{Neural networks seamlessly use the predictive power of fast dynamical fields and slow physical drivers}
\label{sec:con-seamless}
In many predictive statistical approaches aimed at studying weather and climate phenomena, researchers discuss separately the effects of fast dynamical fields and slow physical drivers. For extreme heatwaves, see for instance the interesting works using the analogue method for understanding the effect of fast dynamical drivers~\cite{yiou2014anawege,yiou2019stochastic} and some complementary works on slow drivers~\cite{straaten22}. This dichotomy makes perfect sense given the time scale separation and the complexity of the different approaches. There is however a need for methods that combine both at the same time. For instance, if one wants to quantity the respective impacts of these two types of drivers, one needs a method able to compute predictability skills by dealing with the two types of fields together.  

In this work, we have demonstrated that  neural networks handle, without any practical or methodological difficulty, the 500 hPa geopotential height (fast dynamical field) and soil moisture (slow physical driver). This is in contrast to a method that would explicitly build the effect of averaging over fast drivers, conditioned on the slow drivers, which would require a lot of tricky computations. Moreover, the predictive approach provides actual numbers that quantify the respective role of the two types of fields.

\subsubsection{Probabilistic forecast as a tool for physical analysis of drivers}
\label{sec:con-driver-physical-analysis} 

Current weather models can also handle seamlessly the prediction of the effects of fast dynamical fields and of slow physical drivers. They are actually certainly the most precise way to make such studies. However the objective of using neural network is different and complementary. Neural networks with probabilistic forecast provide a statistical model, which associates to each set of drivers a predictive skill. Alternatively suppressing the different predictors, is a way to estimate the causal relation between any set of fields and the event of interest. This can be used for a posteriori statistical studies, in order to perform fast and efficient process studies, and to analyse the impact of different drivers. Making similar studies with weather or climate model would be extremely difficult in practice and would require huge computations.  We warn however, that any inference about information content from machine learning experiments, assumes that the learning is of a good quality.

We have analyzed, quantitatively, the relative potential of soil moisture and 500-hPa geopotential height, in triggering extreme heatwaves. By adding or removing different fields, and comparing the prediction skills, we can see which are the main drivers. For instance, we have demonstrated that the 2-meter temperature carries part of the predictive information of both the soil moisture or the 500 hPa geopotential height, however we conclude that it carries no new significant information by itself. We have also demonstrated that geopotential height at other altitudes or isopressure levels, carry nearly no new information that can be tapped with a 8,000 long dataset with the given  neural network. 

Those examples illustrate the potential use of  neural network for other process studies in weather and climate dynamics. The key point is the quantitative nature of the analysis. 

\subsection{Conclusions for extreme heatwave drivers}


The main conclusions for extreme heatwave prediction are as follows.The 500 hPa geopotential height combined with soil moist contains the most useful information in the short run, with only a very small improvement of the skill provided by adding the 2m temperature. The prediction skill associated to the 500 hPa geopotential height decays approximately exponentially with a decay time of about 7 days. Soil moisture contains very important complementary information, with a plateau skill that does not decay much on timescales of order of 15 days to a month. This corresponds to the conditional probability to observe extreme heatwaves for some given soil moisture, independently of the dynamics. These two sets of information seem to add up when the two fields are used together, to make the best possible prediction.

We have also concluded that the set of 500-hPa geopotential fields which are selected by the neural network as having a large probability to lead to long-lasting heatwaves, are consistently distributed around a characteristic hemispheric pattern dominated by wavenumber 3  Rossby waves with a shift poleward of the cyclonic anomalies and a shift equatorward of the anticyclonic anomalies. This pattern is also seen in composite maps, conditioned on extreme heatwaves, which are plotted independently of the neural network. An analogous wavenumber 3 pattern has already been observed for European and Scandinavian long-lasting heatwaves~\cite{Ragone18}. This consistency shows that the neural network is either able to recognize this pattern, or is able to recognize other characteristic features of extreme heatwaves which correlate with this pattern. Understanding further those very interesting observations, and the dynamical nature of this pattern, will require to develop the interpretability of machine learning approaches as further discussed in section~\ref{sec:per-lack-of-data}.

\subsection{Key perspectives}

\subsubsection{Perspectives for the lack of data problem for machine learning for climate studies}
\label{sec:per-lack-of-data}

We have concluded in the previous sections that the requested dataset length for convergence of the training of statistical models based on deep architectures is much longer than reanalysis datasets. Since the latter are so short, the use of model data is necessary. But on the other hand, climate models are more biased compared to reality than reanalysis datasets are. There is thus a need to couple the use of climate model and reanalysis datasets in order to make the best of their complementary potential. A natural way is to use transfer learning: first learning from extremely long climate model datasets, and then reusing the weights of the learned model as an initial condition for a new training for the reanalysis dataset. Such a transfer learning approach has already been used in several past works in atmospheric sciences and climate studies~\cite{Ham}.

However this approach (transfer learning) might not be sufficient, for instance if the climate model dataset itself does not contain enough characteristic events. This is probably the case for studying rare or extremely rare events. In order to improve the prediction skill, it is natural to assume that the rare extreme event samples are requested rather than data corresponding to the typical states of the system. Testing this assumption motivates the case for importance sampling algorithms. Regarding the difficulty of sampling exceptionally rare extreme events, e.g. unprecedented heatwaves, we have recently developed rare event simulation techniques that are able to multiply by several orders of magnitude the number of observed heatwaves with PlaSim model~\cite{Ragone18} and with CESM (the NCAR model used for CMIP experiments)~\cite{Ragone21}. We are currently working on coupling these rare event simulations with the machine learning forecast developed in this paper. The point is to improve both rare event simulations using machine learning forecast, and machine learning forecast using the unprecedented heatwave statistics obtained with rare event simulations. We have already coupled machine learning simulations with rare event algorithms, for simple academic models~\cite{Lucente2022}. Coupling the rare event simulations with  neural networks is a very interesting albeit complex perspective to solve the key fundamental issue of lack of data in the science of climate extremes.

Another important perspective is to develop a new  neural framework that would be suited for a better physical interpretability of the prediction skills, in order to increase dynamical and physical understanding. 

In weather and climate dynamics, ensemble forecast by a weather system (for instance ECMWF), is considered as the reference probabilistic medium range forecast. The ensemble members are used to make probabilistic predictions, for instance at medium range or sub-seasonal time scales. An important question for the future will be to compare the skill of machine learning probabilistic forecasts compared to ensemble forecast by weather systems, for instance for predicting extreme heatwaves. 

A key point though is that the two methods, weather systems based on the equations of physics and data assimilation, and learned statistical models, have completely different uses and perspectives, and are highly complementary. On one hand, machine learning  alone is unable to learn the current state of the atmosphere precisely, and to incorporate the wealth of available observation data dealt with by weather systems. A weather system does not make just a single prediction on a specific event, but compute the full state of the system. But on the other hand, a weather system needs dedicated infrastructures and millions of computation hours, while an already trained statistical model usually makes a forecast in less than a second on a laptop. A statistical model can be used to assess the probability from any field, not just the ones in the historical records or the forecasted ones. Then it is extremely likely that weather systems will remain the reference for actual real time medium range forecasts, while statistical model can be used for process studies, driving rare event simulations, statistical analysis, or cheap forecast for targeted scope, or possibly subseasonal to seasonal forecasts.

\subsubsection{Perspectives for extreme heatwaves}
\label{sec:per-heatwaves}


We have argued in section~\ref{sec:con-driver-physical-analysis}  probabilistic forecast issued by neural networks provide a way to estimate the relations between any sets of predictor fields and the event of interest. This is then a tool to quantitatively study the role of each process, very efficiently and practically. Moreover, the quantitative nature of the relation between the predictors and the prediction gives also the opportunity to compare those relations for different models, different datasets, different climates. 

Using this tool opens the door to hundreds of process studies. It could be used for instance to further  ascertain the impact of other slow drivers~\cite{straaten22} on extreme heatwaves and other extreme events, and how they combine with fast dynamical drivers to produce them. One could also use this tool for the purpose of assessing model biases, in order to make climate change studies by comparing different datasets with different climates, and finally to make much more precise impact studies of extreme events. As an important example, it has been demonstrated that local thermodynamics drives monthly midlatitude summertime temperature variance~\cite{vargas2020projected}, and that CMIP model might have some bias in reproducing this effect~\cite{vargas2020projected}. Then, training neural networks on different CMIP models could be used for intercomparison, specifically studying the effect of local thermodynamics on extreme events. 

Given its very easy implementation and its scientific potential, we hope that the deep learning methodology we developed will be used to address many key questions related to extreme events, and other large scale atmosphere and climate phenomena.
\\

\section*{Code and data availability statement}

The coding resources for this work, such as the python and jupyter notebook files, are available on a GitHub page \hypertarget{https://github.com/georgemilosh/Climate-Learning}{https://github.com/georgemilosh/Climate-Learning} branch ``noxarray'' and ``main'' and is part of a larger project at ENS de Lyon with multiple collaborators in the branch ``subm2''. We do not have the infrastructure to make the 8,000 year PlaSim dataset available online at this time, but it might be shared to interested colleagues, whenever feasible in practice. 

\section*{Acknowledgement}

This work was supported by the ANR grant SAMPRACE, project ANR-20-CE01-0008-01 (F. Bouchet). This work has received funding through the ACADEMICS grant of the IDEXLYON, project of the Universit\'e de Lyon, PIA operated by ANR-16-IDEX-0005. We acknowledge CBP IT test platform (ENS de Lyon, France) for ML facilities and GPU devices. The platform operates the SIDUS solution \cite{SIDUS} developed by Emmanuel Quemener. This work was granted access to the HPC resources of CINES under the DARI allocations A0050110575,  A0070110575, A0090110575 and A0110110575 made by GENCI.  We acknowledge the help of Alessandro Lovo in maintaining the GitHub page.


\bibliography{references}



\end{document}